\documentclass[useAMS,usenatbib]{mn2e}
\usepackage{graphicx}
\title[New Galactic symbiotic stars with SALT]{Identification of new Galactic symbiotic stars with SALT. I. Initial discoveries and other emission line objects\thanks{Based on observations made with the Southern African Large Telescope (SALT) under programme 2013-1-RSA\_POL-001.}}
\author[Miszalski and Miko{\l}ajewska]{Brent Miszalski$^{1,2}$\thanks{E-mail: brent@saao.ac.za} and Joanna Miko{\l}ajewska$^{3}$\\
$^{1}$South African Astronomical Observatory, PO Box 9, Observatory, 7935, South Africa\\
$^{2}$Southern African Large Telescope Foundation, PO Box 9, Observatory, 7935, South Africa\\
$^{3}$Nicolaus Copernicus Astronomical Centre, Bartycka 18, 00716 Warsaw, Poland\\
}
\begin{document}

\date{Accepted . Received ; in original form }

\maketitle
\begin{abstract}
   We introduce the first results from an ongoing, systematic survey for new symbiotic stars selected from the AAO/UKST SuperCOSMOS H$\alpha$ Survey (SHS). The survey aims to identify and characterise the fainter population of symbiotic stars underrepresented in extant catalogues. The accreting white dwarfs (WDs) in symbiotic stars, fuelled by their red giant donors with high mass loss rate winds, make them promising candidates for type Ia supernovae. Several candidates were observed spectroscopically with the Southern African Large Telescope (SALT). A total of 12 bona-fide and 3 possible symbiotic stars were identified. The most remarkable example is a rare carbon-rich symbiotic star that displays coronal [Fe~X] $\lambda$6375 emission, suggesting it may be a supersoft X-ray source with a massive WD. Several other emission line objects with near-infrared colours similar to symbiotic stars are listed in an appendix, including 6 B[e] stars, 4 planetary nebulae (PNe), 2 possible Be stars, one [WC9] Wolf-Rayet (WR) central star of a PN and one WC9 WR star. These initial discoveries will help shape and refine the candidate selection criteria that we expect will uncover several more symbiotic stars as the survey progresses. 

\end{abstract}


\section{Introduction}
Symbiotic stars are interacting binaries with the longest orbital periods. A rich variety of phenomena are created in the interplay between the high mass loss rate wind of the red giant and its hot, accreting white dwarf companion (see e.g. Miko{\l}ajewska 2012). There are less than 300 Galactic symbiotics known (e.g. Allen 1984; Kenyon 1986; Miko{\l}ajewska, Acker \& Stenholm 1997; Belczy\'nski et al. 2000; Corradi et al. 2008, 2010a; Miszalski, Miko{\l}ajewska \& Udalski 2013, hereafter MMU13), considerably fewer than population estimates of $3\times10^3$ (Allen 1984), $3\times10^4$ (Kenyon et al. 1993) or 3--4 $\times10^5$ (Munari \& Renzini 1992; Magrini et al. 2003). To better estimate the true size of the population and the connection with type Ia supernovae (see e.g. Di Stefano 2010; Dilday et al. 2012), we require a more representative population of symbiotic stars. New surveys must be undertaken to find and characterise the full extent of the fainter population of symbiotic stars that are underrepresented in current catalogues (e.g. Belczy\'nski et al. 2000). 

The strong H$\alpha$ emission of symbiotic stars ensures that they stand out in large H$\alpha$ surveys such as the INT Photometric H$\alpha$ Survey (IPHAS; Drew et al. 2005) and the AAO/UKST SuperCOSMOS H$\alpha$ Survey (SHS; Parker et al. 2005). The surveys constitute a promising discovery medium for new symbiotic stars with their excellent sensitivity to faint H$\alpha$ emission line stars and large sky coverage. Nineteen new symbiotic stars have so far been spectroscopically confirmed from the IPHAS survey (Corradi et al. 2008, 2010a,b, 2011; Corradi \& Giammanco 2010; Corradi 2012; Rodr\'iguez-Flores et al. in preparation). Their target selection strategy combined selections in the IPHAS colour-colour plane, to select for H$\alpha$ emission, as well as the near-infrared (NIR) Two Micron All Sky Survey (2MASS, Skrutskie et al. 2006) colour-colour plane, to select for red giants. With the SHS, MMU13 visually selected H$\alpha$ emitters across $\sim$35 deg$^2$ towards the Galactic Bulge and spectroscopically confirmed 20 new symbiotic stars, as well as several other possible symbiotic star candidates. These new IPHAS and SHS discoveries represent a typically fainter population than the known sample of symbiotic stars (see e.g. figure 16 of MMU13). 

No systematic survey to increase the number of southern Galactic symbiotic stars has yet been made using the SHS catalogue photometry. The SHS contains catalogue photometry in three filters, an H$\alpha$ interference filter, a Short-Red ($SR$) filter (5900--6900 \AA) that serves as an off-band for H$\alpha$, and an $I$-band filter (see Parker et al. 2005 for details). Pierce (2005) investigated the position of Belczy\'nski et al. (2000) symbiotic stars in the SHS colour-colour plane ($SR-I$ and $H\alpha-SR$), but did not conduct a search for new symbiotic stars. Miszalski et al. (2008) demonstrated SHS catalogue photometry could be used to find H$\alpha$ emitters, i.e. compact PNe, forming the bulk of the MASH-II PN catalogue. The focus of MASH-II was, however, to identify new PNe rather than symbiotic stars, resulting in a bias against H$\alpha$ emitters with the NIR colours of red giants. Such symbiotic star candidates were unfortunately not recorded as part of the MASH-II survey. 

In this paper we introduce an ongoing, systematic survey for new southern Galactic symbiotic stars and present the first discoveries made from spectroscopic follow-up with the Southern African Large Telescope (SALT). Section \ref{sec:sel} outlines the candidate selection from SHS and 2MASS catalogue photometry and Sect. \ref{sec:obs} describes the SALT spectroscopy. Section \ref{sec:new} introduces 12 new symbiotic stars and we conclude in Sect. \ref{sec:conclusion}. Appendix \ref{sec:other} presents several other H$\alpha$ emission line objects.

\section{Candidate selection} 
\label{sec:sel}
During the first phase of this project relatively loose criteria were chosen to allow for several types of H$\alpha$ emission line stars to be observed. These early non-symbiotic contaminants will inform stricter selection criteria and will help determine the completeness of the survey during later phases. We therefore postpone a full discussion of the symbiotic star candidate selection criteria until the spectroscopic follow-up of candidates is more advanced. Nevertheless, we can outline here the candidate selection process during the first exploratory phase. Catalogue photometry was retrieved from the SHS website\footnote{http://www-wfau.roe.ac.uk/sss/halpha/haobj.html} for H$\alpha\le15$ mag and SHS field centres with RA$\ga$14 h. A variety of colour-colour and magnitude cuts were applied to the catalogue photometry using the \textsc{stilts}\footnote{http://www.star.bris.ac.uk/$\sim$mbt/stilts/} package of command-line tools for processing table data (Taylor 2006, 2011). The cuts were designed to incorporate the colours and magnitudes of MMU13 symbiotic stars that we have assumed to be repesentative of the relatively faint and reddened symbiotic star population that awaits discovery elsewhere in the SHS. The resultant candidates were then cross-matched against 2MASS using the Centre de Donn\'ees astronomiques de Strasbourg (CDS) cross-match service.\footnote{http://cdsxmatch.u-strasbg.fr/xmatch} Further constraints were applied to the 2MASS photometry and duplicates were removed using the 2MASS ID as the unique name for each object. Candidates were then visually inspected using colour-composite images as described in Miszalski et al. (2008), with promising candidates selected based on their similar appearance to MMU13 symbiotic stars.

\section{SALT RSS spectroscopy}
\label{sec:obs}
We observed several candidates spectroscopically with the Robert Stobie Spectrograph (RSS; Burgh et al. 2003; Kobulnicky et al. 2003) on the queue-scheduled Southern African Large Telescope (SALT; Buckley, Swart \& Meiring 2006; O'Donoghue et al. 2006) under programme 2013-1-RSA\_POL-001 (PI: Miszalski). A single RSS configuration was adopted with the PG900 grating and a 1.5 arcsec wide slit at a position angle of 0$^\circ$ to give wavelength coverage from $\sim$4340--7420 \AA\ at a resolution of 6.2 \AA\ (full-width at half maximum, FWHM) and a reciprocal dispersion of 0.97 \AA\ pixel$^{-1}$. Two CCD chip gaps are present in RSS spectra, approximately covering 5355--5415\AA\ and 6400--6460\AA, with the exact coverage and wavelengths depending on spectrograph flexure at the time of observation. Table \ref{tab:obs} records the 2MASS point source catalogue designation, date and exposure time of the observations presented in this paper. The exposure time given corresponds to the long exposure that is taken after a short 30 or 60 s exposure to measure H$\alpha$ unsaturated. A xenon arc lamp reference spectrum is taken after the long exposure, while no flat-field frames are taken. Other candidates were also observed, but these will be included in future papers to help define the candidate selection criteria. Relatively long exposure times were selected for most candidates to correspond with the relaxed observing conditions of the programme, namely any lunar phase, thin cloud and seeing up to 3.0--3.5 arcsec. The observations were therefore taken under a wide range of conditions and sometimes in considerably better seeing. Three objects were observed during October when SALT visibility windows were rapidly shrinking, necessitating very short exposure times. This meant two objects were observed well before astronomical twilight, namely 2MASS J17422035$-$2401162 and 2MASS J18272892$-$1555547, observed 25 and 33 minutes before twilight, respectively.

A higher priority weighting was given to objects with redder 2MASS colours in an effort to address the difficulties in discovering D-type symbiotic stars (Corradi et al. 2008, 2010a; MMU13). In D-type (dusty) symbiotic stars the red giant is a Mira variable (Whitelock 1987) that is often completely obscured by dust at optical wavelengths. To confirm a symbiotic star the red giant must be observed and in D-types the high extinction makes this task consistently difficult when only optical spectroscopy is available. 

Basic reductions were applied using the \textsc{pysalt} package (Crawford et al. 2010) and cosmic ray events were cleaned using the \textsc{lacosmic} package (van Dokkum 2001). Wavelength calibration was performed using the standard \textsc{iraf} tasks \textsc{identify}, \textsc{reidentify}, \textsc{fitcoords} and \textsc{transform}, and one-dimensional spectra were extracted using the \textsc{iraf} task \textsc{apall}. A relative flux calibration was applied to the extracted spectra using a spectrum of the spectrophotometric standard star LTT 6248 (Hamuy et al. 1994). As the moving pupil design of SALT does not allow for an accurate absolute flux scale, we have normalised the spectra by the average continuum value measured in the region 6200--6300 \AA. The spectra are presented in subsequent sections of the paper and have not been dereddened. The relatively short exposures of 2MASS J17463311$-$2419558 and 2MASS J17422035$-$2401162 that we present later have been lightly smoothed with a boxcar filter.

\begin{table}
   \centering
   \caption{Log of SALT RSS observations. Only the long exposure time is given (see text).}
   \label{tab:obs}
   \begin{tabular}{llr}
      \hline
      Name & Date      & Exposure   \\
       (2MASS J)       & (dd/mm/yy)& (s)        \\
      \hline
      14031865$-$5809349 & 18/06/13 & 1500  \\
      14135896$-$6709206 & 25/05/13 & 1800  \\
      15431767$-$5857221 & 20/05/13 & 1200  \\
      15460752$-$6258042 & 11/05/13 & 1500  \\
      16003761$-$4835228 & 14/05/13 & 1200  \\
      16092019$-$5536100 & 29/04/13 & 1800  \\
      16141537$-$5146036 & 13/05/13 & 1800  \\
      16293215$-$5215338 & 12/06/13 & 1800  \\
      16422739$-$4133105 & 02/05/13 & 1200  \\
      16490162$-$3827431 & 13/07/13 & 1800  \\
      16503229$-$4742288 & 23/05/13 & 1680  \\
      17050868$-$4849122 & 18/09/13 & 900  \\
      17060193$-$3848359 & 29/05/13 & 1500  \\
      17145509$-$3933117 & 17/06/13 & 1500 \\
      17334728$-$2719266 & 10/09/13 & 600  \\
      17391715$-$3546593 & 12/07/13 & 1500  \\
      17400382$-$4019271 & 21/07/13 & 1209  \\
      17422035$-$2401162 & 16/10/13 & 297   \\
      17460199$-$3303085 & 21/09/13 & 1200  \\
      17463311$-$2419558 & 18/10/13 & 500   \\
      18114984$-$2017513 & 24/07/13 & 1500  \\
      18131474$-$1007218 & 22/07/13 & 1250  \\
      18143298$-$1825148 & 22/07/13 & 1200  \\
      18144673$-$2035178 & 25/07/13 & 1800  \\
      18272892$-$1555547 & 22/10/13 & 670 \\
      18285064$-$2432017 & 25/07/13 & 1500  \\
      18300636$-$1940315 & 21/09/13 & 1050  \\
      18332967$-$0333531 & 23/04/13 & 1800  \\
      18340300$-$1132321 & 20/06/13 & 1800  \\
      18481892$+$0143066 & 27/04/13 & 1500  \\
      18482874$-$1237434 & 13/07/13 & 1800  \\
      18500902$-$0515041 & 25/04/13 & 1500  \\
      19003482$-$0211579 & 29/04/13 & 1800  \\
      \hline
   \end{tabular}
\end{table}

\section{New symbiotic stars}
\label{sec:new}
\begin{figure*}
   \begin{center}
\includegraphics[scale=0.55,angle=270]{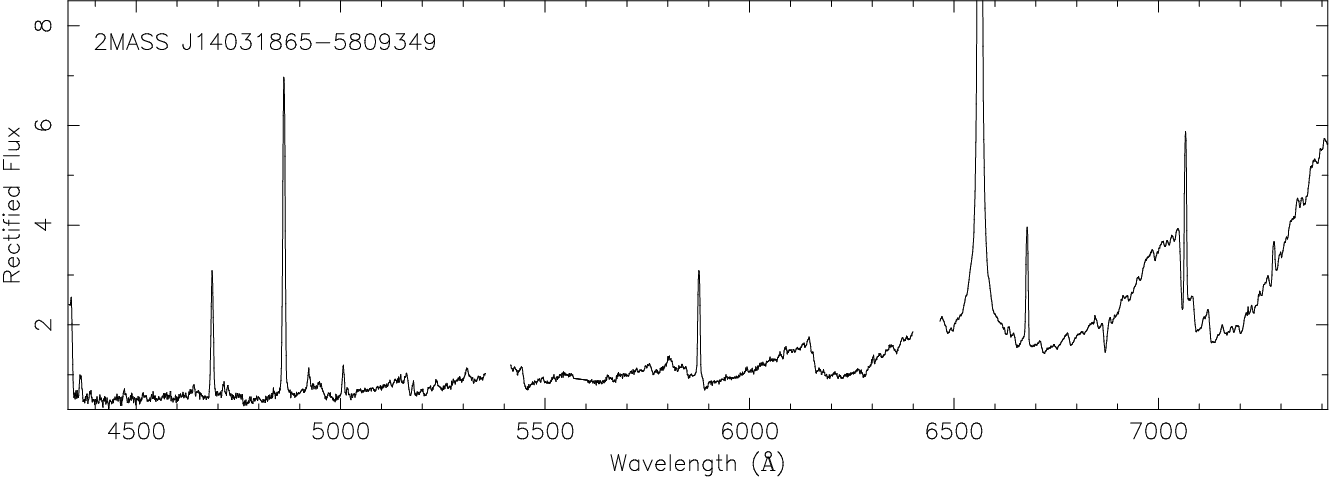}
\includegraphics[scale=0.55,angle=270]{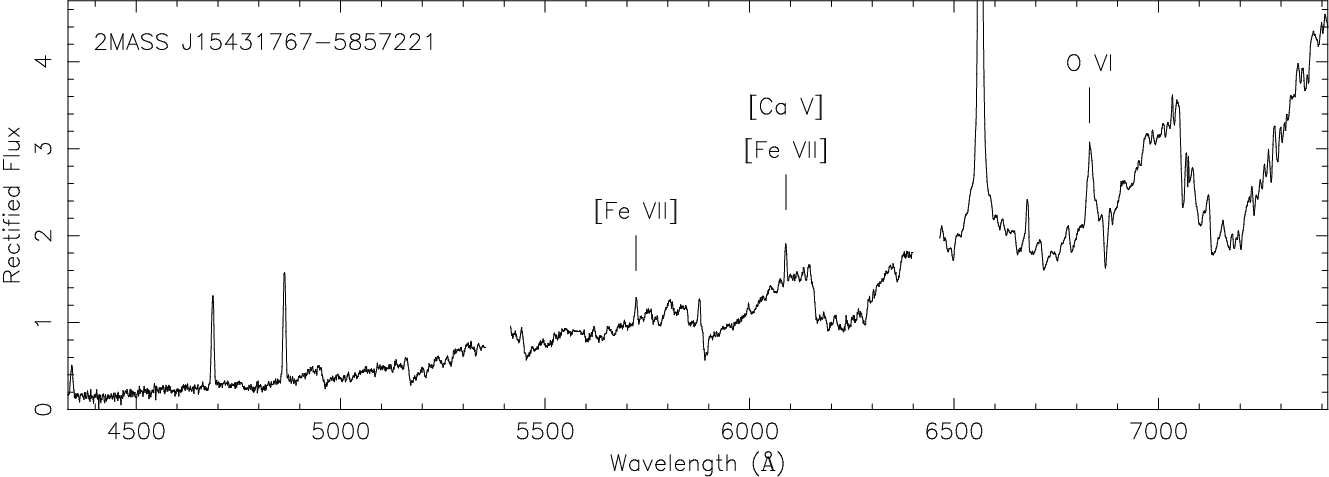}
\includegraphics[scale=0.55,angle=270]{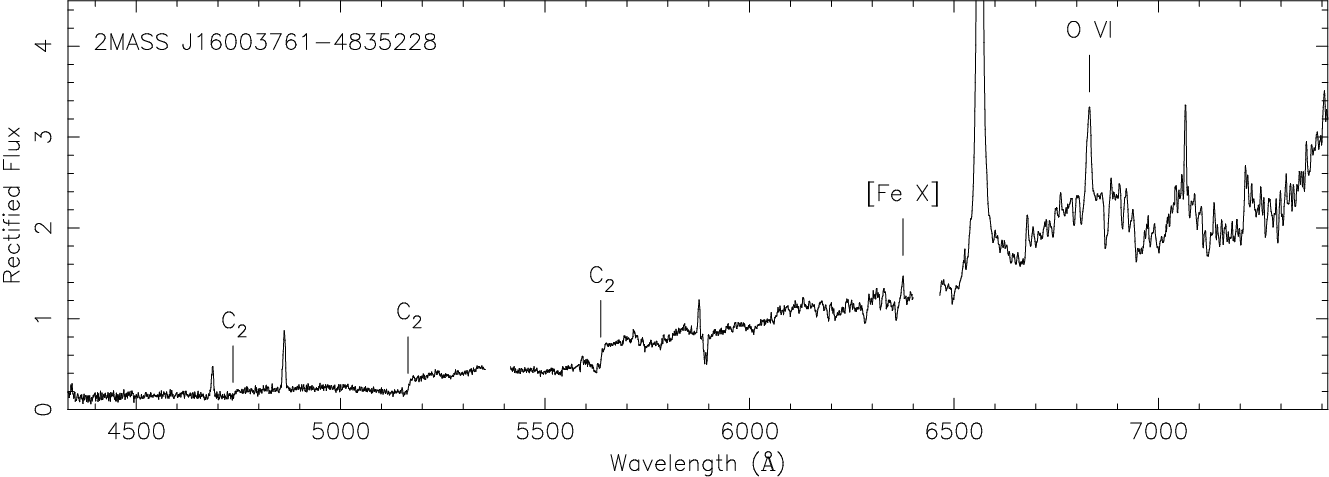}
\includegraphics[scale=0.55,angle=270]{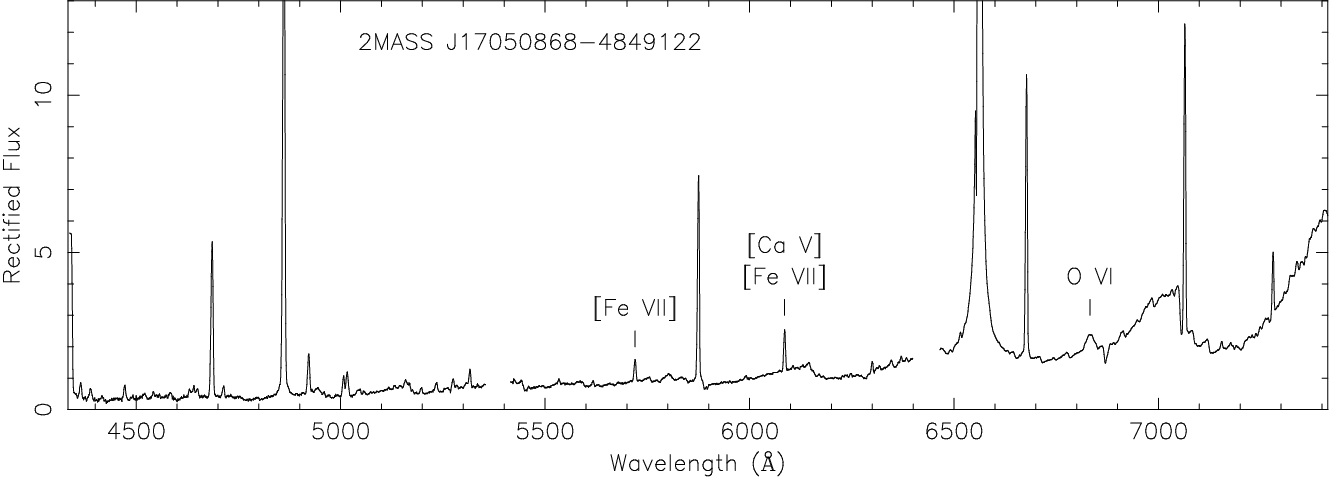}
\includegraphics[scale=0.55,angle=270]{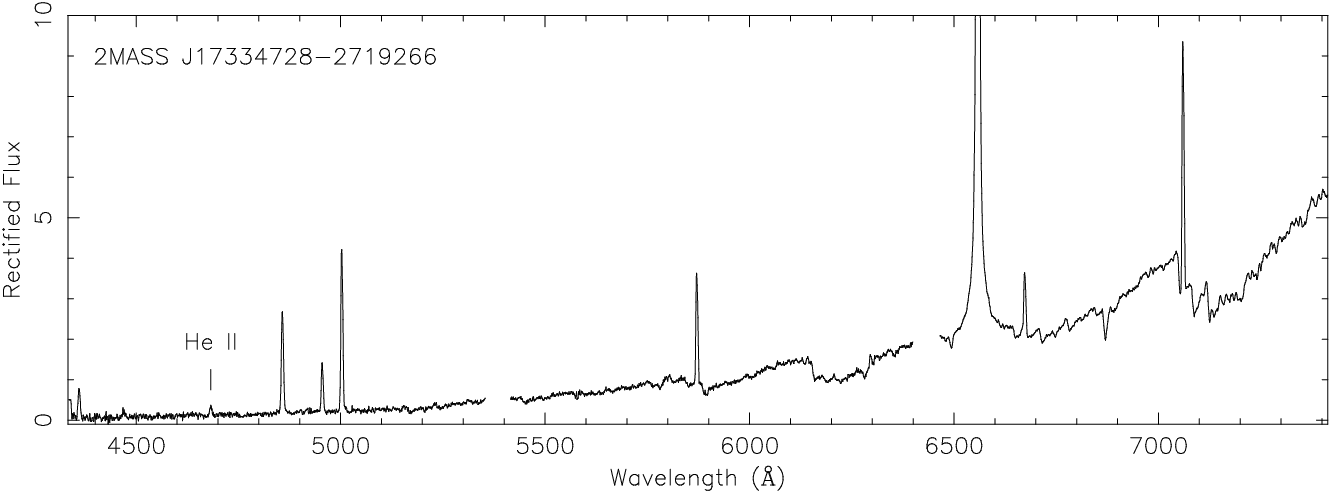}
   \end{center}
   \caption{SALT RSS spectra of new S-type symbiotic stars.}
   \label{fig:s1}
\end{figure*}

\begin{figure*}
   \begin{center}
\includegraphics[scale=0.55,angle=270]{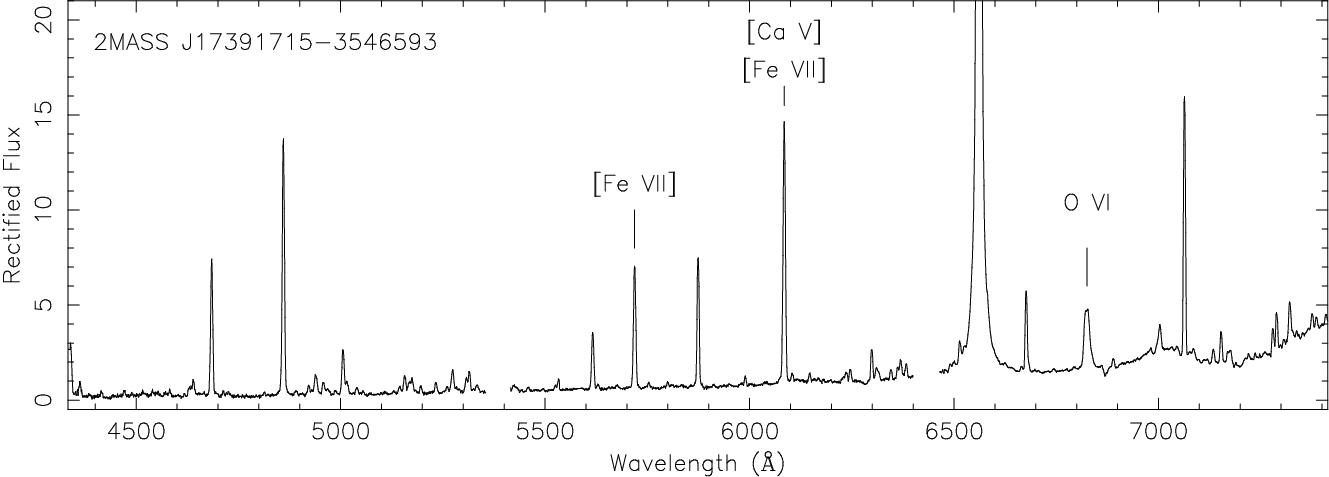}
\includegraphics[scale=0.55,angle=270]{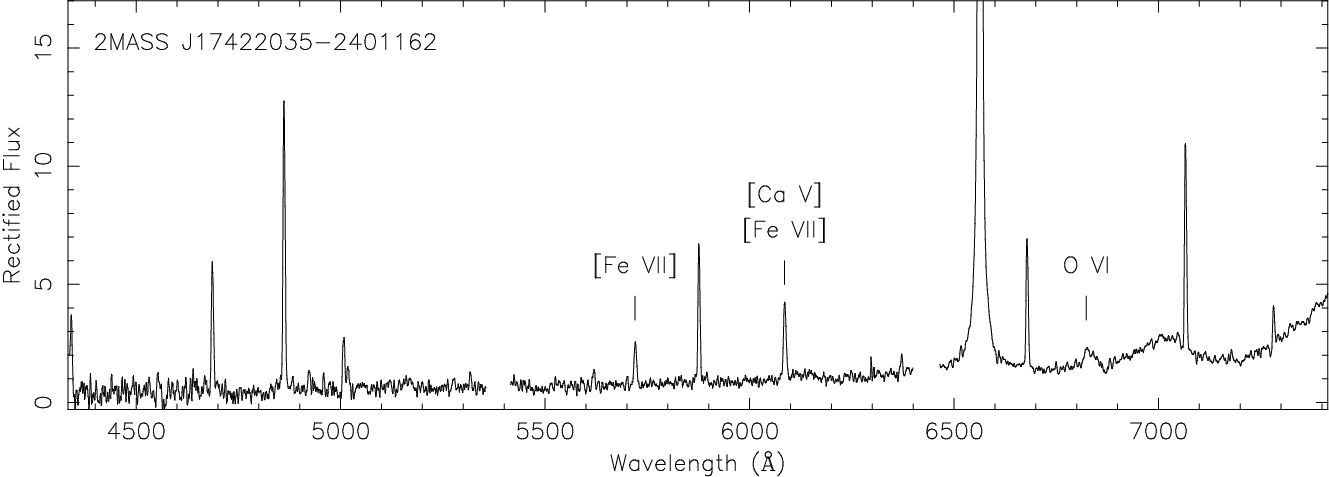}
\includegraphics[scale=0.55,angle=270]{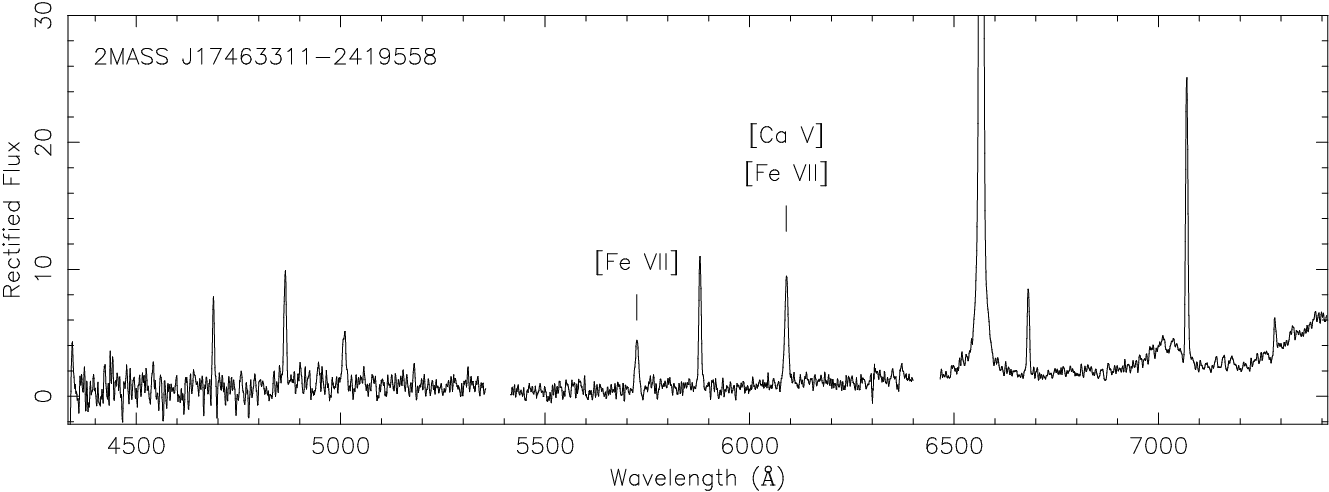}
\includegraphics[scale=0.55,angle=270]{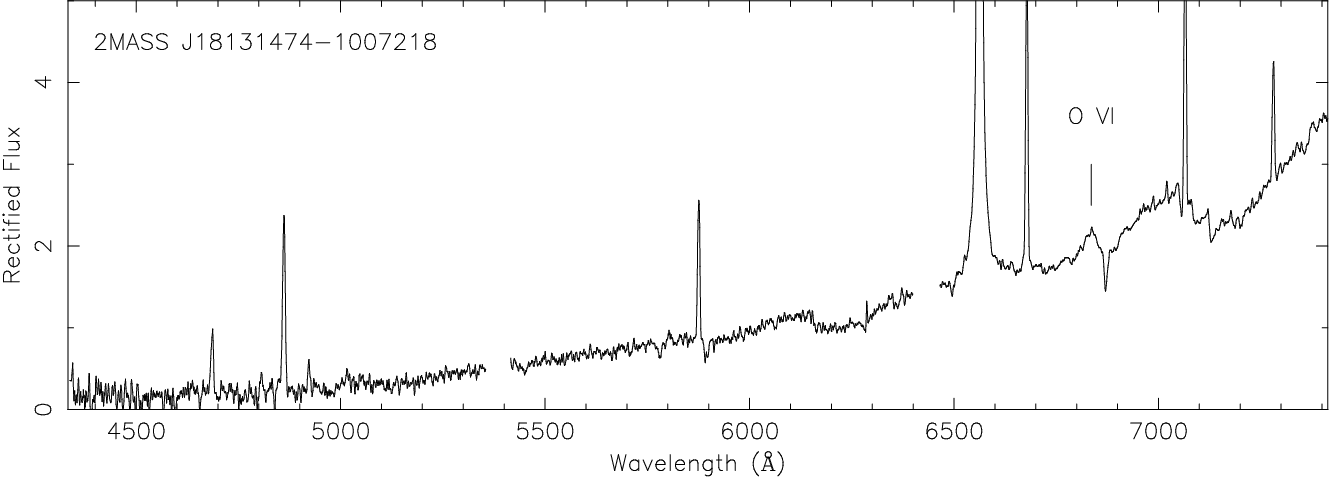}
\includegraphics[scale=0.55,angle=270]{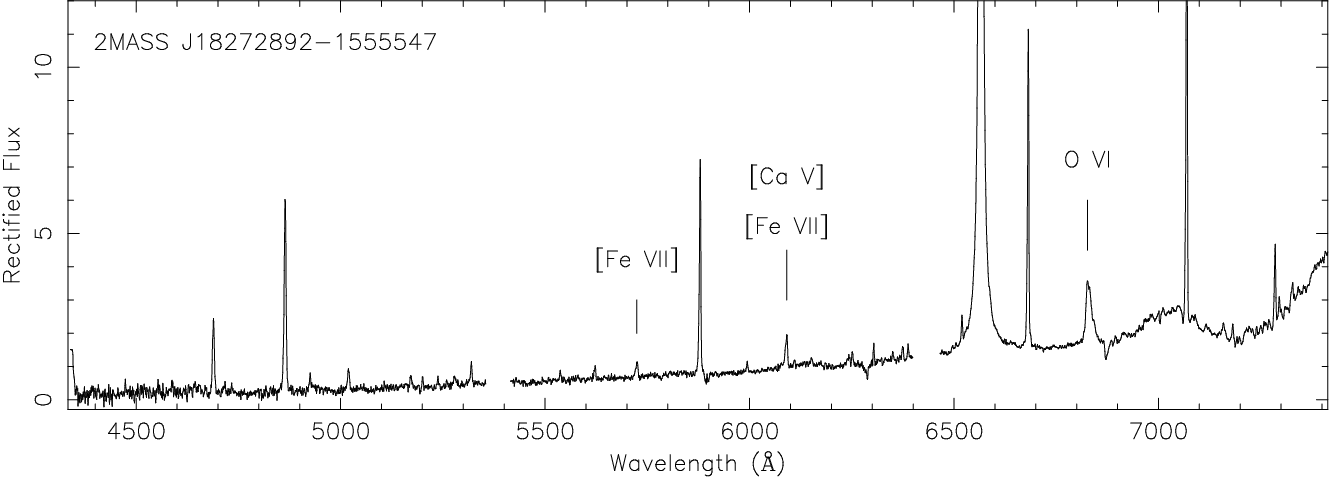}
   \end{center}
   \caption{SALT RSS spectra of new S-type symbiotic stars (continued).}
   \label{fig:s2}
\end{figure*}

\begin{figure*}
   \begin{center}
\includegraphics[scale=0.55,angle=270]{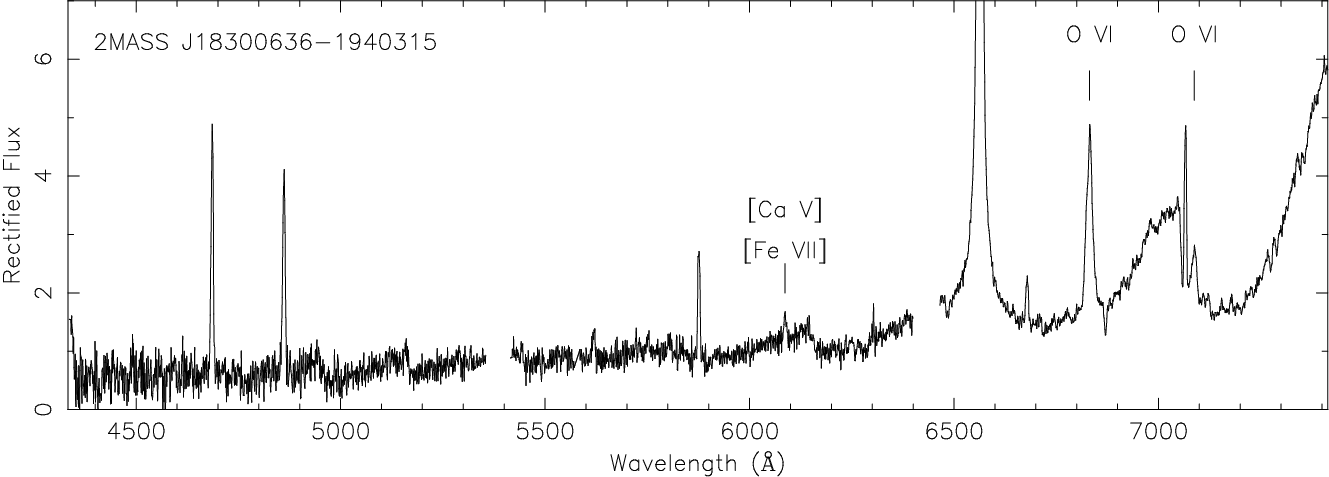}
\includegraphics[scale=0.55,angle=270]{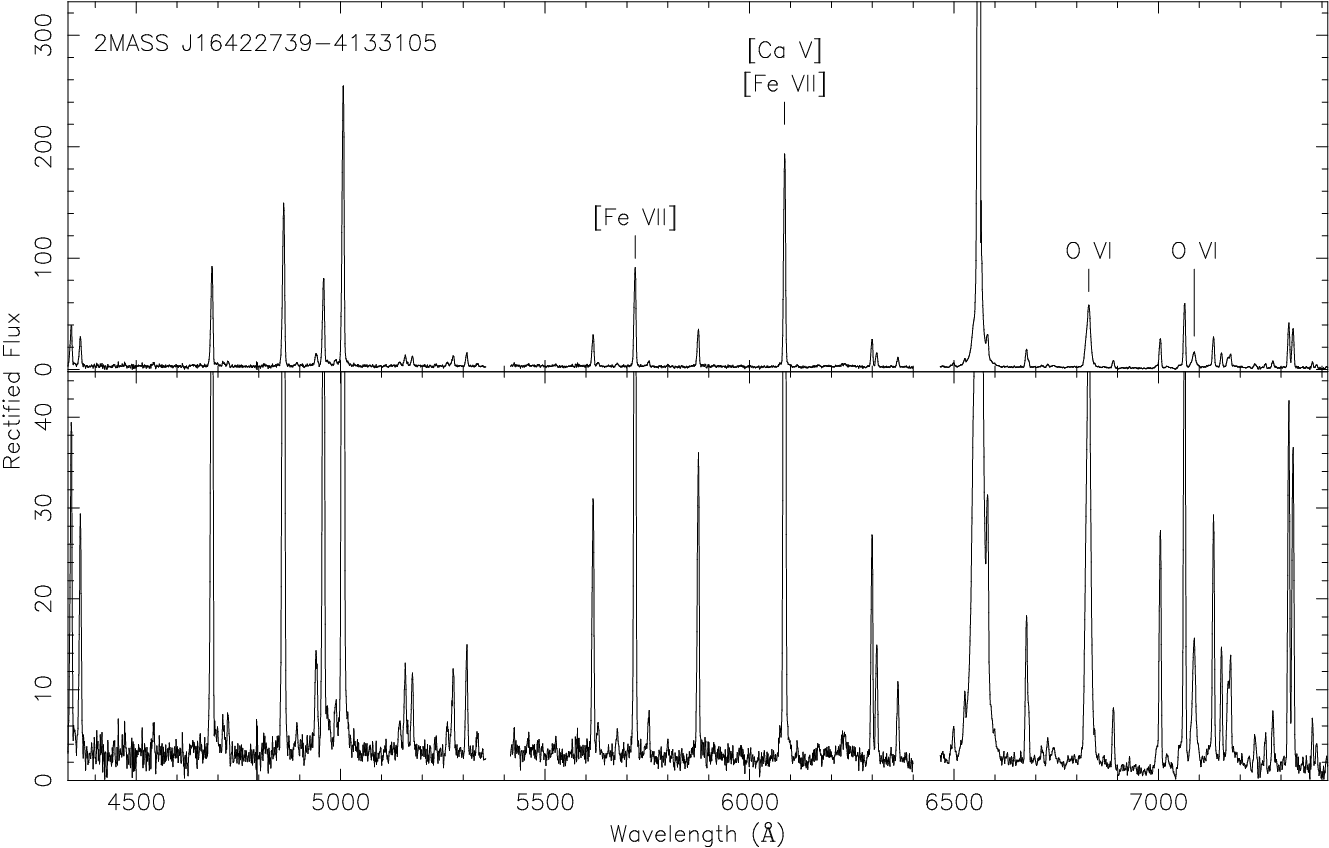}
   \end{center}
   \caption{SALT RSS spectra of the new symbiotic stars 2MASS J18300636$-$1940315 (S-type) and 2MASS J16422739$-$4133105 (D-type).}
   \label{fig:s3}
\end{figure*}

\begin{figure*}
   \begin{center}
\includegraphics[scale=0.55,angle=270]{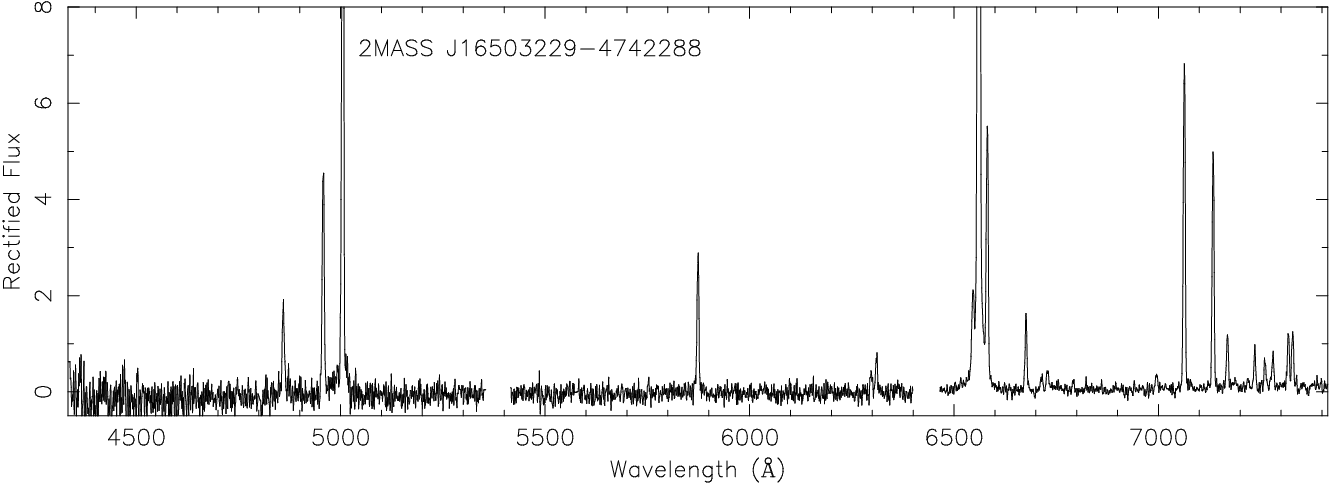}
\includegraphics[scale=0.55,angle=270]{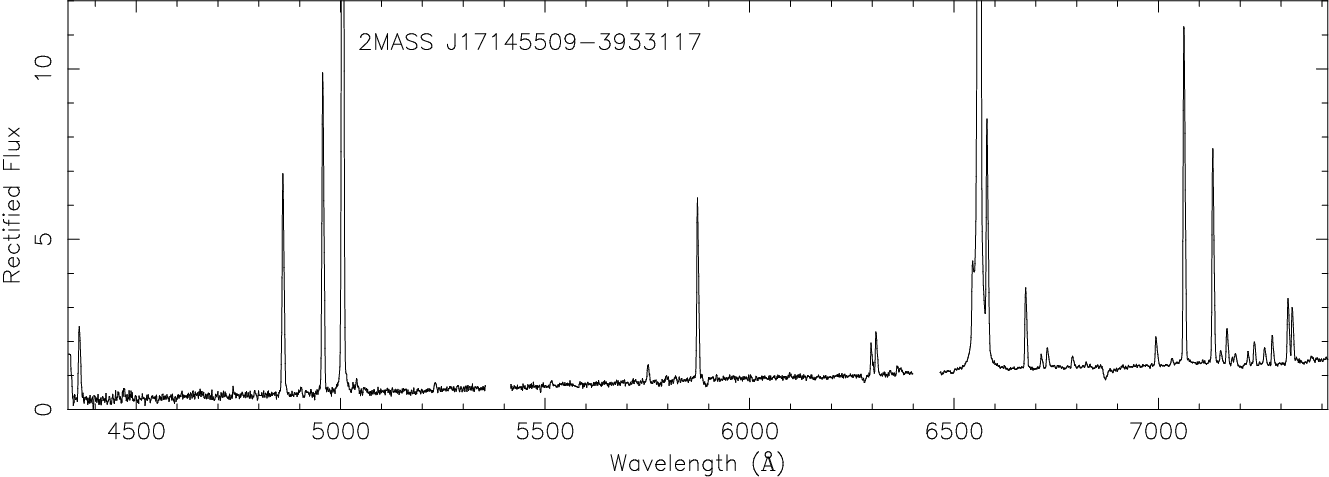}
\includegraphics[scale=0.55,angle=270]{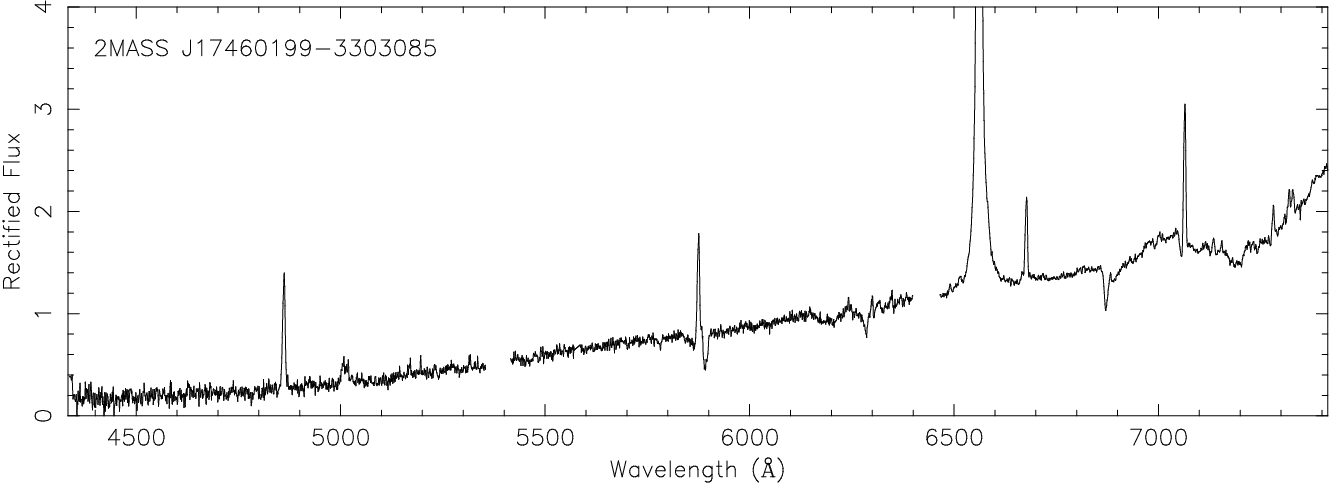}
   \end{center}
   \caption{SALT RSS spectra of the possible symbiotic stars 2MASS J16503229$-$4742288 (D-type), 2MASS J17145509$-$3933117 (S-type) and 2MASS J17460199$-$3303085 (S-type).}
   \label{fig:d1}
\end{figure*}

We have identified 12 bona-fide and three possible symbiotic stars following the classification criteria for symbiotic stars in Belczy\'nski et al. (2000) and MMU13. Their basic properties are presented in Table \ref{tab:new} and their spectra in Figures \ref{fig:s1}, \ref{fig:s2}, \ref{fig:s3} and \ref{fig:d1}. The bona-fide sample consists of 11 S-types and 1 D-type, all of which have multiple necessary features to be classified as symbiotic stars. Strong and broad H$\alpha$ emission and He~I emission lines are found in all objects. The cool components are present in all S-types and their spectral types in Tab. \ref{tab:new} were mostly determined as in MMU13, using the indices of Kenyon \& Fernandez-Castro (1987), except for 2MASS J16003761$-$4835228 in which case we used the Barnbaum, Stone \& Keenan (1996) atlas of carbon stars. The Barnbaum et al. (1996) atlas is based on the revised MK classification of red carbon stars (Keenan 1993). Spectral types of red giants with under-exposed stellar continuua are less certain and are marked with an additional ':'. The red giant in the D-type 2MASS J16422739$-$4133105 was not detected and is likely too obscured by dust to be visible at optical wavelengths. Hot components are readily detected via key emission lines including He~II $\lambda$4686 (all objects), the Raman-scattered O~VI emission lines (Schmid 1989) that are a telltale diagnostic feature of symbiotic stars (8 out of 11 S-types and 1 D-type), and [Fe~VII] $\lambda\lambda$5721, 6089 (7 out of 11 S-types and 1 D-type). The weakest He~II $\lambda$4686 emission is found in 2MASS J17334728$-$2719266, while it is exceptionally strong in 2MASS J18300636$-$1940315. The highest ionization emission line detected is coronal [Fe X] $\lambda$6375 in the carbon-rich 2MASS J16003761$-$4835228, strongly suggesting it may be a supersoft x-ray source (see Sect. \ref{sec:carbon}). 

\begin{table*}
   \centering
   \caption{Basic properties of the new and possible symbiotic stars.}
   \label{tab:new}
   \begin{tabular}{lrrllrrrrrr}
      Name (2MASS J) & $\ell$ ($^\circ$) & $b$ ($^\circ$) & IR type & Spectral type & $J$ & $J-H$ & $H-K_s$ & H$\alpha$ & H$\alpha-SR$ & $SR-I$\\
      \hline
      14031865$-$5809349 & 312.3148 & 3.3967        & S & M4   &10.42&	1.10	&0.38	&12.10	&$-$1.60	&1.44           \\ 
      15431767$-$5857221 & 323.5413 & $-$3.1423     & S & M2.5 &11.22&	1.09	&0.41	&12.43	&$-$1.40	&0.34         \\ 
      16003761$-$4835228 & 332.0679 & 3.2823        & S & C-N5 C$_2$4.5 &10.80&	1.33	&0.58	&13.22	&$-$1.75	&1.89         \\ 
      16422739$-$4133105 & 342.2640 & 3.0318        & D & -    &10.33&	1.37	&1.30	&10.71	&$-$3.30	&$-$1.21         \\ 
      17050868$-$4849122 & 339.1468 & $-$4.6492     & S & M4   &10.04&	1.09	&0.38	&11.81	&$-$2.07	& 2.28         \\
      17334728$-$2719266 & 359.9791 & 3.0663        & S & M2   &9.32	&  1.41	&0.65	& 11.23	&$-$1.79	& $-$0.58      \\
      17391715$-$3546593 & 353.4730 & $-$2.4679     & S & M1.5 &9.81	&  1.37	&0.81	&12.93	&$-$2.60	&0.86         \\
      17422035$-$2401162 & 3.8062   & 3.1975        & S & M2:  &10.20	&  1.21	&0.56	& 12.70	& $-$1.98& 1.05 \\
      17463311$-$2419558 & 4.0423   & 2.2155        & S & M4:  &9.86	& 1.35	& 0.60 &	11.46 &	$-$2.28	& $-$0.86\\
      18131474$-$1007218 & 19.5540  & 3.7375        & S & M0   &10.94 &	1.30	&0.46	&13.75	&$-$1.51	&1.16     \\
      18272892$-$1555547 & 16.0601 & $-$2.0558      & S & M1   &9.16  &	1.36	&0.62	&11.14	&$-$2.95	& 3.67\\
      18300636$-$1940315 &  13.0226 & $-$4.3378     & S & M3.5 &11.07	&  1.11	&0.41	&12.56	&$-$2.21	& 1.23	     \\
                         &          &               &   &      &     &        &     &        &        &     \\     
      16503229$-$4742288 & 338.5095 & $-$2.0533     & D?& -    &13.99&	2.44	&2.07	&14.54	&$-$2.78	&$-$0.73       \\
      17145509$-$3933117 & 347.6561 & $-$0.5638 & D?& -    &11.92&	2.30	&1.75	&13.21	&$-$2.10	&$-$0.14 \\ 
      17460199$-$3303085 & 356.5301 & $-$2.2173     & S?& K5-M0&9.86	&  1.16	&0.53&	9.50	&$-$2.34	& 0.81         \\
      \hline
   \end{tabular}
\end{table*}

Three possible symbiotic stars were identified that almost meet the classification criteria. In the case of the possible D-types 2MASS J16503229$-$4742288 and 2MASS J17145509$-$3933117, a PN-like spectrum is observed to coincide with red 2MASS colours typical of D-type symbiotic stars ($J-H$ and $H-K_s$ are both $\ga$2.0), however a red giant is not seen in the optical spectra and further NIR observations are required to determine if one is present. If a significant amount of the reddening towards 2MASS J16503229$-$4742288, 2MASS J16422739$-$4133105 and 2MASS J17145509$-$3933117 is circumstellar in origin, then their Mira components would not be visible in the optical spectrum. In known D-type symbiotic stars, an average intrinsic colour for unobscured symbiotic Miras of $\langle(J-K)_0\rangle=1.45$ mag can be adopted from the period-colour relation (Gromadzki et al. 2009). Our new D-type systems have observed $J-K_s$ colours of 2.67, 4.51 and 4.05 mag for 2MASS J16422739$-$4133105, 2MASS J16503229$-$4742288 and 2MASS J17145509$-$3933117, respectively, corresponding to $A_K\sim0.8$, $\sim$2.0 and $\sim$1.8 mag (Cardelli et al. 1989). Assuming Case B conditions, we measure the reddening from the nebular Balmer decrements to be $A_V=6.0$, 8.0 and 7.1 mag, respectively, corresponding to $A_K=0.69$, 0.91 and 0.81 (Cardelli et al. 1989). The reddening calculated in this way is a upper limit and may be smaller due to optical depth effects. As these values are less than the expected $A_K$ values, a significant amount of the reddening should indeed be circumstellar, explaining the absence of the cool components in our spectra. Similar outcomes appear in many other D-type symbiotics (Miko{\l}ajewska et al. 1999; MMU13). Another alternative explanation for the two possible D-types might be a PN with a dusty Wolf-Rayet central star (e.g. Sect. \ref{sec:wr}), but these typically have temperatures of $\sim$30 kK and are unable to ionise the [Ar~IV] and [Ar~V] emission lines observed. These lines are more common in D-type symbiotic stars than in PNe and are evident in several objects in MMU13.

\subsection{Notes on some individual objects}
We searched for matches to all new and possile symbiotic stars in SIMBAD and VizieR. Most objects are either not catalogued or have minimal associated information. As would be expected, the only IRAS detections in our sample are found in the D-type 2MASS J16422739$-$4133105 (IRAS 16389$-$4737) and possible D-types 2MASS J16503229$-$4742288 (IRAS 16468$-$4737) and 2MASS J17145509$-$3933117 (IRAS 17114$-$3929). In the following we provide some information on some noteworthy individual objects. 

\subsubsection{2MASS J14031865$-$5809349}
Skiff (2013) lists this object as emission line object Wray 15-1167, referenced in Stephenson \& Sanduleak (1977b), but it does not appear in Stephenson \& Sanduleak (1977a). A marginal detection of [Fe~VII] emission lines may also be present in the spectrum, however the strong He~II $\lambda$4686 leaves no doubt that the hot component is present. 

\subsubsection{2MASS J16003761$-$4835228 - a carbon-rich supersoft X-ray source candidate}
\label{sec:carbon}
The new S-type 2MASS J16003761$-$4835228 is remarkable because of its C-N5 C$_2$4.5 carbon star in the Barnbaum et al. (1996) scheme, making it the latest in a list of rare carbon-rich Galactic symbiotic stars after IPHAS J205836.43$+$503307.2 (Corradi et al. 2011) and H1-45 (MMU13). A total of five are now known, including SS~38 and AS~210 (Gromadzki et al. 2009). These discoveries seem to appear with greater frequency in the relatively fainter H$\alpha$ selected candidates from which they were found compared to those in Belczy\'nski et al. (2000). It may be explained by the more representative nature of the reddened or more distant sample. 

The most exceptional property of 2MASS J16003761$-$4835228, however, is the presence of the weak [Fe~X] $\lambda$6375 \AA\ emission line, that combined with the fairly strong Raman scattered O~VI emission band, places 2MASS J16003761$-$4835228 amongst those symbiotic stars showing the highest degree of ionization. The coronal [Fe~X] $\lambda$6375 line appears in classical novae (e.g. McLaughlin 1953), symbiotic novae (e.g. PU Vul, Andrillat \& Houziaux 1994), and symbiotic recurrent novae (Williams et al. 1991) during their nebular phase, however it is exceedingly rare in classical symbiotic stars. In principle, the presence of [Fe~X] is well-documented only in SMC~3, which is also the brightest supersoft X-ray source (SSXS) associated with a symbiotic binary (e.g. Jordan et al. 1996; Orio et al. 2007). Kato et al. (2013) argued that in SMC~3 both the soft X-rays and  the coronal [Fe~X] line originate from a hot and dense ($n_e \sim 10^9$--$10^{10}\, \rm cm^{-3}$) nebula around the nuclear-burning massive white dwarf which scatters X-rays from the white dwarf.

There are only a handful SSXS (with all photons detected below 1.0 keV) among known symbiotic stars (M\"urset et al. 1997; Luna et al. 2013), however [Fe~X] has not been reported for any of them except SMC~3 (although we cannot exclude that a weak line could be overlooked). All of them have been found in metal-poor populations, either at high Galactic latitudes (e.g. AG Dra and StHa32), or in the SMC (SMC~3 and Lin~358). The most likely origin of the soft X-rays in these systems is quasi-steady H-shell burning on the white dwarf (e.g. Miko{\l}ajewska et al. 1995; Orio et al. 2007). On the other hand, most of the known symbiotic stars contain such thermonuclear shell-burning white dwarfs and the scarcity of SSXS among them indicates that there must be efficient ways to quench X-ray emission. One obvious way to suppress this emission is interstellar extinction which affects practically all known Galactic systems. However, even unreddened symbiotic white dwarfs are embedded in the wind of their cool giant companion and they may also emit their own wind. Nielsen et al. (2013) showed that soft X-ray emission from steadily accreting and nuclear-burning white dwarfs can be very efficiently attenuated even for modest circumbinary mass loss rates. The final effect will also depend strongly on the metallicity as it determines the neutral and ionized gas opacity (Nielsen et al. 2013).

The presence of a cool, carbon-rich and possibly metal-poor giant in 2MASS J16003761$-$4835228, together with the very high degree of ionization indicated by [Fe~X], makes it a very appealing candidate for a SSXS. However, detecting soft X-rays in 2MASS J16003761$-$4835228 may be severely hindered or blocked entirely by neutral gas and dust (to a lesser extent) in the Galactic plane at the  $b=3.03^\circ$ latitude of 2MASS J16003761$-$4835228. The high reddening towards 2MASS J16003761$-$4835228 is evident in the high $E(B-V)\sim1.6$ mag from the Galactic extinction map of Schlafly \& Finkbeiner (2011).

\subsubsection{2MASS J17334728$-$2719266} 
The S-type is included in Terzan \& Gosset (1991) as variable star V 2513 that reached $R=8.0$ mag, however the authors note this upper limit to be uncertain. This is an increase of up to 5.0 mag over the SHS $SR$ magnitude (Tab. \ref{tab:new}), which could be explained either by a Z-And outburst (typically 1--3 mag, Miko{\l}ajewska 2003) or a symbiotic nova outburst (several magnitudes). 
\subsubsection{2MASS J17391715$-$3546593} 
The S-type was discovered first as PN K 5-8 (Kohoutek 1994), with the remark of 'stellar (low excitation)'. Escudero \& Costa (2001) later remarked that the spectrum of K 5-8 does not seem to be that of a PN and it was therefore not included in their analysis. 

\subsubsection{2MASS J17422035$-$2401162} 
\begin{figure*}
   \begin{center}
      \includegraphics[scale=1.00,angle=270]{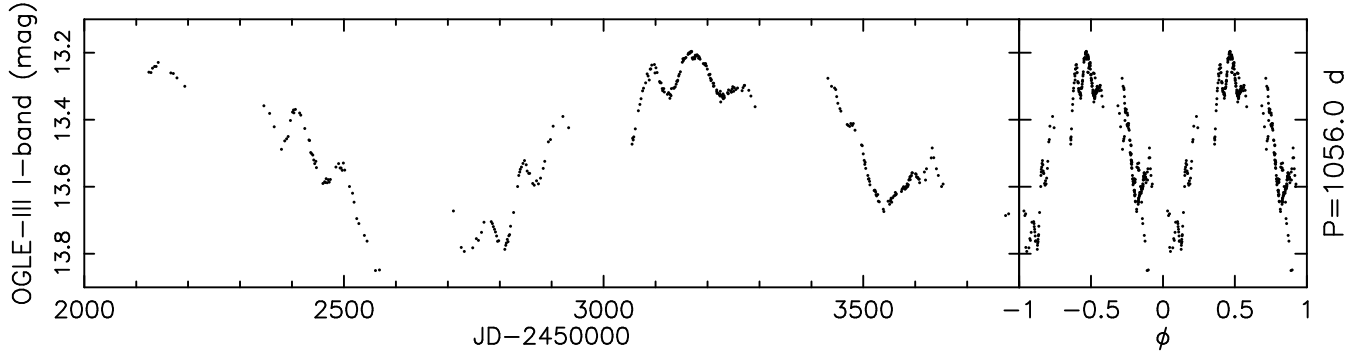}
   \end{center}
   \caption{The OGLE-III $I$-band light curve of the new S-type symbiotic star 2MASS J17422035$-$2401162 (Soszy{\'n}ski et al. 2013).}
   \label{fig:lc}
\end{figure*}

The S-type is classified as a semi-regular variable in the OGLE-III Galactic bulge long period variable (LPV) catalogue with a primary period of 1056 d (Soszy{\'n}ski et al. 2013). Figure \ref{fig:lc} reproduces the $I$-band light curve where we have used an ephemeris at light curve minimum of 2676.4 d (JD$-$2450000) + $1056E$. The light curve is consistent with those of other S-type symbiotic stars where the orbital period is modulated by pulsations from the red giant (e.g. MMU13; Gromadzki, Miko{\l}ajewska \& Soszy\'nski 2013). The average $I$-band magnitude is 13.45 mag, while we determine a less certain average $V$-band magnitude of 17.40$\pm$0.25 mag from only three OGLE-III $V$-band measurements. 

\subsubsection{2MASS J17460199$-$3303085}
The possible S-type was catalogued as KW 055 in the H$\alpha$ emission star catalogue of Kohoutek \& Wehmeyer (2003). 
The spectrum resembles an S-type symbiotic, but we lack the crucial features needed to confirm the presence of a hot component. Features such as He~II $\lambda$4686 may be suppressed below the noise level of the spectrum due to the high reddening towards the object. Additional spectra may reveal these features at a different orbital phase, as observed in 002.86-01.88 with O~VI emission only appearing in one out of two spectra (MMU13).

\section{Conclusions}
\label{sec:conclusion}
We have presented the first results from an ongoing survey for new symbiotic stars in the southern Galactic plane. Candidates were selected from SHS and 2MASS photometry catalogues with relatively loose selection criteria based on the observed properties of MMU13 symbiotic stars. Several new symbiotic stars were confirmed spectroscopically with SALT, in addition to many other H$\alpha$ emission line objects that will refine our selection criteria as the survey progresses. We expect to discover many more symbiotic stars in the near future, as we observe candidates selected from all SHS fields, most of which will be found towards the Galactic bulge.

The main conclusions are as follows: 
\begin{itemize}
   \item Twelve new symbiotic stars were identified across a range of Galactic longitudes. The sample consists of 11 S-types and 1 D-type, of which 8 objects show the telltale Raman scattered O~VI emission band (Schmid 1989). 
   \item Most remarkable in the sample is 2MASS J16003761$-$4835228, a carbon-rich symbiotic star that exhibits a weak coronal [Fe~X] $\lambda$6375 emission line. This suggests 2MASS J16003761$-$4835228 may be a supersoft X-ray source, produced by quasi-steady H-shell burning onto the WD (e.g. Miko{\l}ajewska et al. 1995; Orio et al. 2007). Such systems are prime candidates to identify massive accreting WDs that form promising type Ia supernovae (e.g. Kato et al. 2013). If the supersoft X-rays are not entirely attentuated (e.g. Nielsen et al. 2013), their detection may prove very challenging due to the high interstellar absorption towards 2MASS J16003761$-$4835228. The C-N5 C$_2$4.5 carbon star component makes it the fifth known carbon-rich Galactic symbiotic star known, following other recent carbon-rich discoveries by Corradi et al. (2011) and MMU13. There may be a propensity for fainter symbiotic star populations to exhibit a greater frequency of carbon-rich symbiotic stars. We speculate that this may be explained by more reddened or more distant symbiotic stars included in our sample and further survey work is required to test this hypothesis. 
   \item The discovery of D-type symbiotic stars proved unexpectedly more difficult than in MMU13. Despite efforts to bias the spectroscopic follow-up towards objects with dusty NIR colours, only one D-type and two possible D-types were identified. The most numerous contaminating objects included six B[e] stars and reddened PNe. Dusty [WC] Wolf-Rayet central stars of PNe also occupy this parameter space, as evidenced by the discovery of a [WC9] nucleus in K3-18, previously suspected of being a symbiotic star (Giammanco et al. 2011). A dusty [WC] central star can be excluded if [Ar~IV] and [Ar~V] emission lines are present, since many cool [WC] central stars do not ionise [O~III] which has a lower ionisation potential. 
\end{itemize}

\section{Acknowledgements}
All of the observations reported in this paper were obtained with the Southern African Large Telescope (SALT) and we would like to thank the Polish SALT time allocation committee for their generous award of SALT time. JM is supported by the Polish National Science Center grant number DEC-2011/01/B/ST9/06145.
This research made use of the cross-match service, the SIMBAD database and the VizieR catalogue access tool, all provided by CDS, Strasbourg, France. This research has made use of SAOImage DS9, developed by Smithsonian Astrophysical Observatory, and we would also like to thank Mark Taylor for developing and maintaining the \textsc{stilts} and \textsc{topcat} software packages used heavily in this work. This publication makes use of data products from the Two Micron All Sky Survey, which is a joint project of the University of Massachusetts and the Infrared Processing and Analysis Center/California Institute of Technology, funded by the National Aeronautics and Space Administration and the National Science Foundation.

\appendix
\clearpage
\newpage
\section[]{Other emission line objects}
\label{sec:other}
\begin{figure*}
   \begin{center}
      \includegraphics[scale=0.7,angle=270]{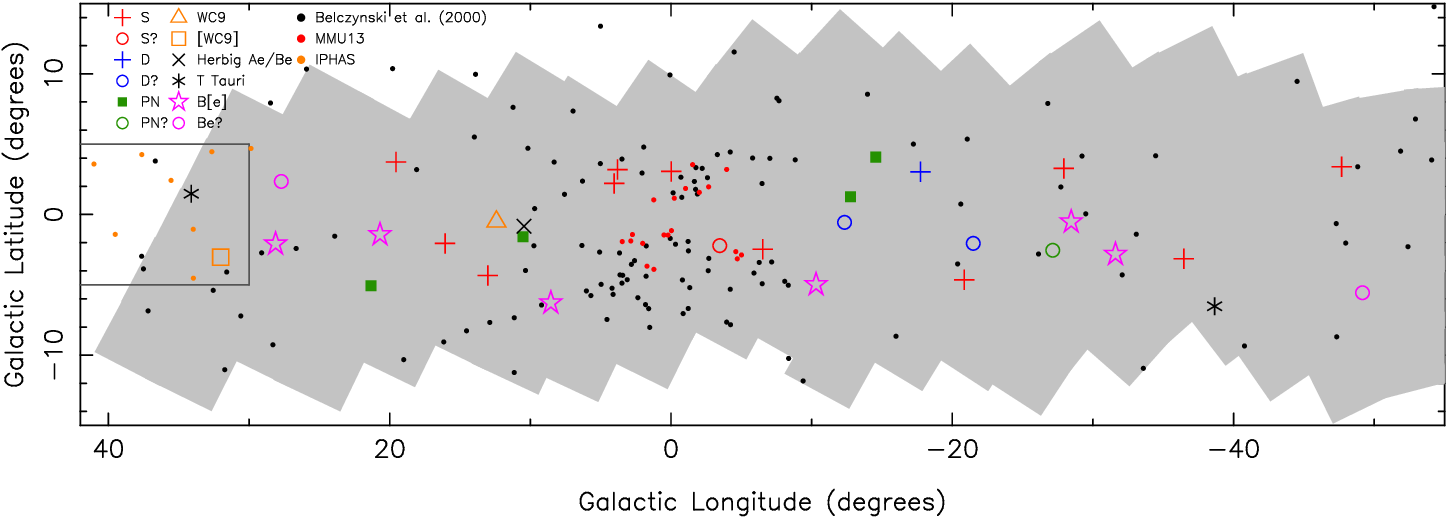}
   \end{center}
   \caption{Galactic distribution of our sample. Small dots show symbiotic stars from Belczy\'nski et al. (2000), MMU13 and IPHAS (Corradi 2012 and ref. therein). Survey footprints of SHS and IPHAS are indicated by the grey shaded region and solid lines, respectively.}
   \label{fig:gal}
\end{figure*}

\begin{figure}
   \begin{center}
      \includegraphics[scale=0.5,angle=270]{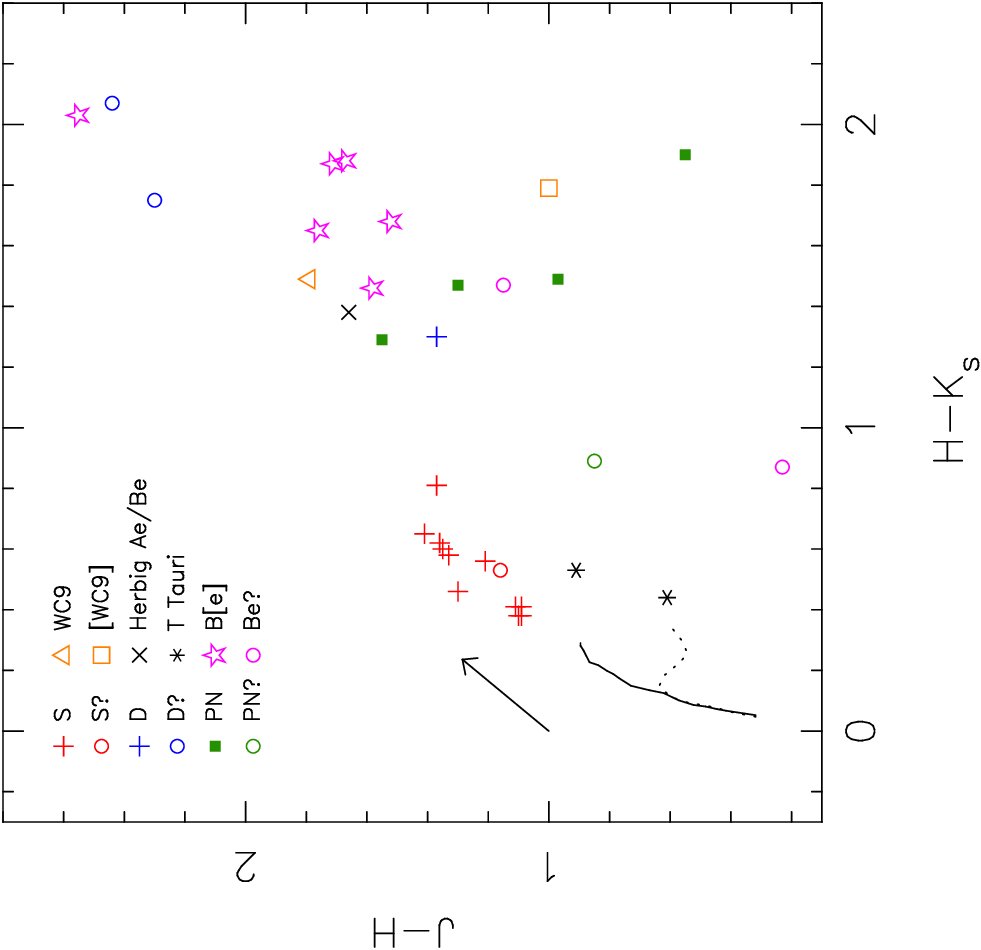}
   \end{center}
   \caption{The location of our sample in the 2MASS colour-colour plane. The arrow indicates the reddening vector for $A_V=3.1$ mag (Cardelli et al. 1989). The unreddened main-sequence and giant sequence stellar locii, obtained from Bessell \& Brett (1988) and transformed into the 2MASS system using Carpenter (2001), are displayed as dotted and solid black lines, respectively.}
   \label{fig:2mass}
\end{figure}

Included in Tab. \ref{tab:obs} are different H$\alpha$ emission line stars or nebulae of various kinds. As previous studies have found (e.g. Corradi et al. 2008, 2010a; MMU13), their NIR colours often overlap D-type symbiotics, and to a lesser extent, S-type symbiotics. Table \ref{tab:other} collates their basic properties and we display the Galactic distribution of the entire sample in Fig. \ref{fig:gal}. Classifications were made based on features in the SALT RSS spectra and their positions in the 2MASS colour-colour diagram (Fig. \ref{fig:2mass}) that provides additional diagnostic power (e.g. Schmeja \& Kimeswenger 2001; Miko{\l}ajewska 2004; Corradi et al. 2008).

\begin{table*}
   \centering
   \caption{Basic properties of other emission line objects.}
   \label{tab:other}
   \begin{tabular}{lrrlrrrrrr}
      Name (2MASS J) & $\ell$ ($^\circ$) & $b$ ($^\circ$) & Type  & $J$ & $J-H$ & $H-K_s$ & H$\alpha$ & H$\alpha-SR$ & $SR-I$\\
      \hline
      16293215$-$5215338 & 332.8628 & $-$2.5470 & PN? & 14.50&0.85&0.89&13.86&$-$2.27&	0.09 \\
      16490162$-$3827431 & 345.4383 &4.0800 & PN & 14.02 &	1.55 &	1.29 &	14.31 &	$-$2.26 &	$-$0.54\\
      17060193$-$3848359 & 347.2329 & 1.2633 & PN & 14.06&0.55&1.90&12.30&$-$3.19	&$-$1.57\\
      18144673$-$2035178 & 10.5302  & $-$1.5843 & PN & 14.24&	1.30&	1.47&	14.02	&$-$2.50&	1.11\\ 
      18482874$-$1237434 & 21.3298 & $-$5.0762 & HDC PN & 14.28 &	0.97 &	1.49	& 12.11	& $-$2.23 &	$-$1.52\\
                         &          &           &      &      &       &          &       &       &          \\     
      16092019$-$5536100 & 328.3956 & $-$2.8404 & B[e] & 14.26&	1.52&	1.68&	12.50	&$-$2.53&	$-$0.63\\ 
      16141537$-$5146036 & 331.5484 & $-$0.5366 & B[e] & 12.80&	1.76&	1.65&	13.99	&$-$2.47&	1.23\\
      17400382$-$4019271 & 349.6922 & $-$5.0003 & B[e] & 13.10&	1.71&	1.87&	11.54	&$-$2.04&	0.54  \\
      18285064$-$2432017 & 8.5468   & $-$6.2951 & B[e] & 13.24&	2.55&	2.03&	11.85	&$-$2.65&	$-$0.37\\
      18340300$-$1132321 & 20.6919  & $-$1.4408 & B[e] & 12.28&	1.58&	1.46&	12.99	&$-$1.68&	1.01\\
      18500902$-$0515041 & 28.1099  & $-$2.1063 & B[e] & 11.61&	1.67&	1.88&	11.88	&$-$1.42&	0.26\\
                         &          &           &      &      &       &     &       &       &          \\     
      14135896$-$6709206 & 310.8334 & $-$5.5589 & Be? & 13.77&	0.23&	0.87&	11.87	&$-$1.71&	$-$0.14 \\
      15460752$-$6258042 & 321.3460 & $-$6.5227 & T Tauri & 11.25 &	0.61 & 0.44 & 13.18 &	$-$1.48 &	1.67 \\
      18114984$-$2017513 & 10.4547  & $-$0.8388 & Herbig Ae/Be & 11.81&	1.66	&1.38&	13.65 &	$-$1.81&	2.52 \\
      18143298$-$1825148 & 12.4094  & $-$0.5022 & WC9 & 9.04 &	1.79&	1.49&	11.18&	$-$1.72&	2.23          \\
      18332967$-$0333531 & 27.7103  & 2.3568 & Be?  & 14.15&	1.15&	1.47&	13.39	&$-$2.38	&$-$1.21\\
      18481892$+$0143066 & 34.1058  & 1.4781            & T Tauri&  12.85&	0.91&	0.53	&14.36&	$-$1.57&	0.58             \\ 
      19003482$-$0211579 & 32.0148  & $-$3.0362 & [WC9] & 13.29&1.00&1.79&12.60&$-$2.05&$-$0.05        \\ 
      \hline
   \end{tabular}
\end{table*}

\subsection[]{B[e] and possible Be stars}
\begin{figure*}
   \begin{center}
\includegraphics[scale=0.55,angle=270]{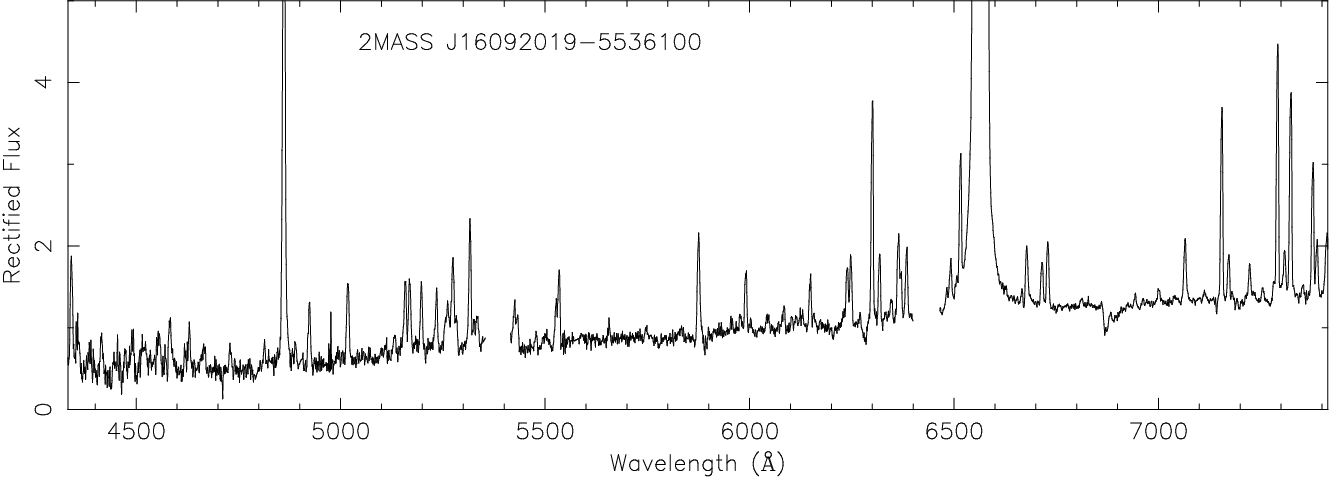}
\includegraphics[scale=0.55,angle=270]{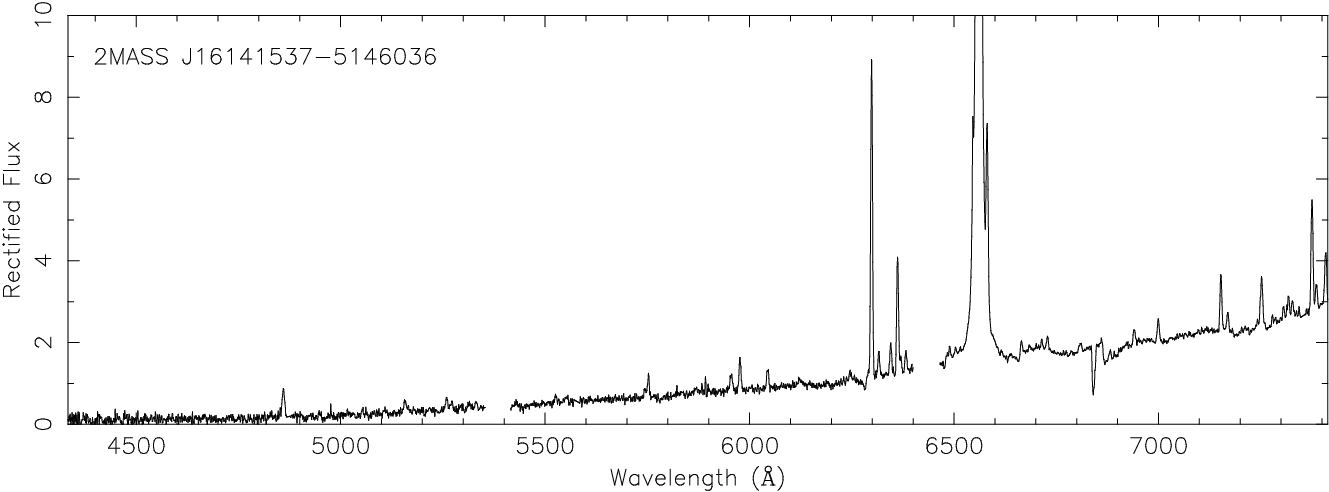}
\includegraphics[scale=0.55,angle=270]{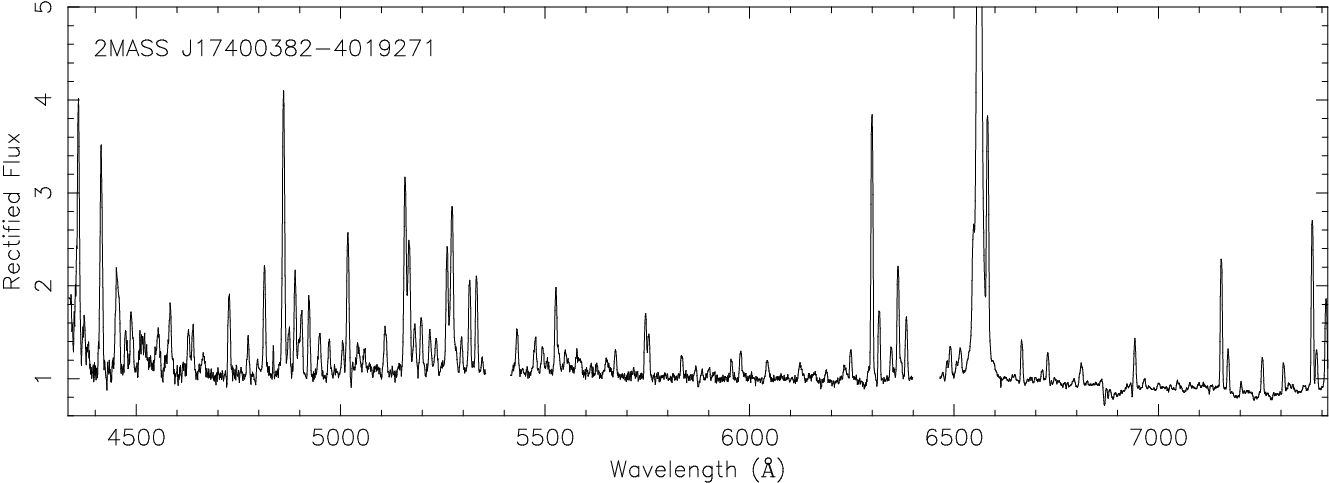}
\includegraphics[scale=0.55,angle=270]{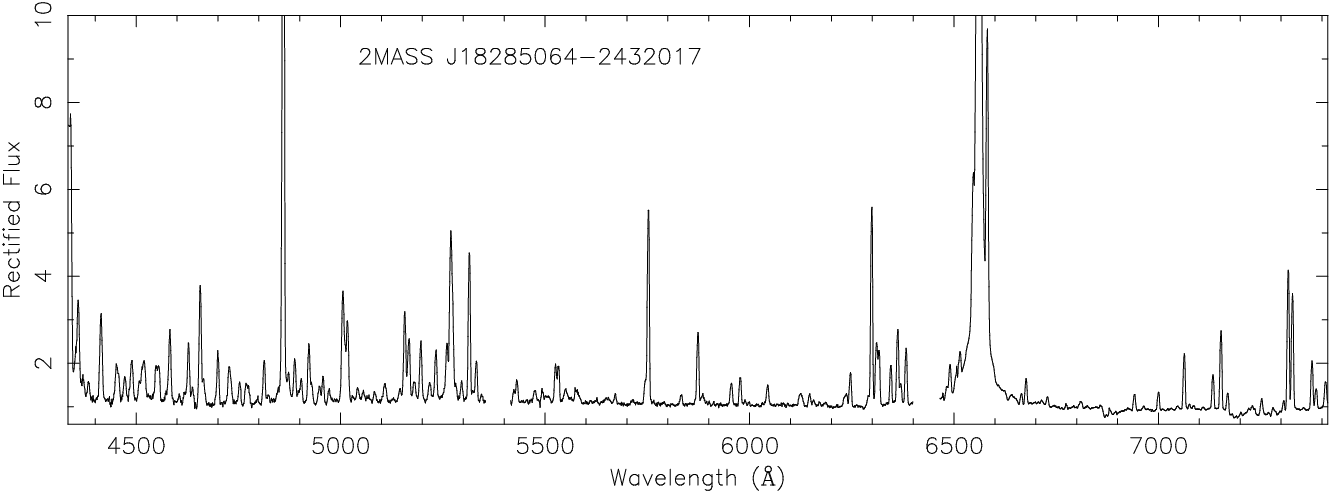}
\includegraphics[scale=0.55,angle=270]{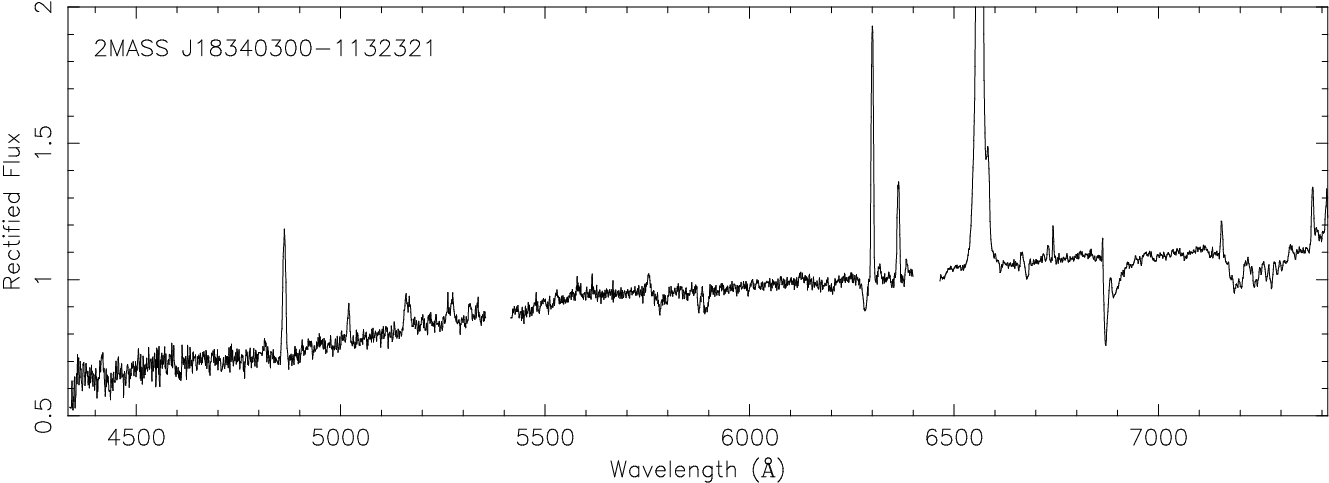}
   \end{center}
   \caption{SALT RSS spectra of B[e] stars.}
   \label{fig:be1}
\end{figure*}

\begin{figure*}
   \begin{center}
\includegraphics[scale=0.55,angle=270]{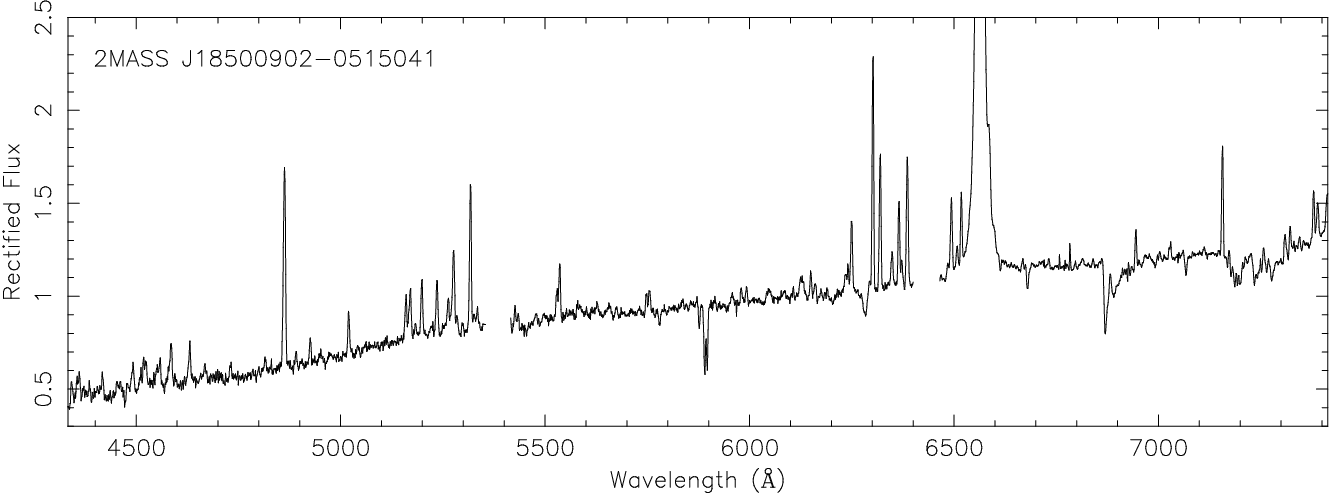}
\includegraphics[scale=0.55,angle=270]{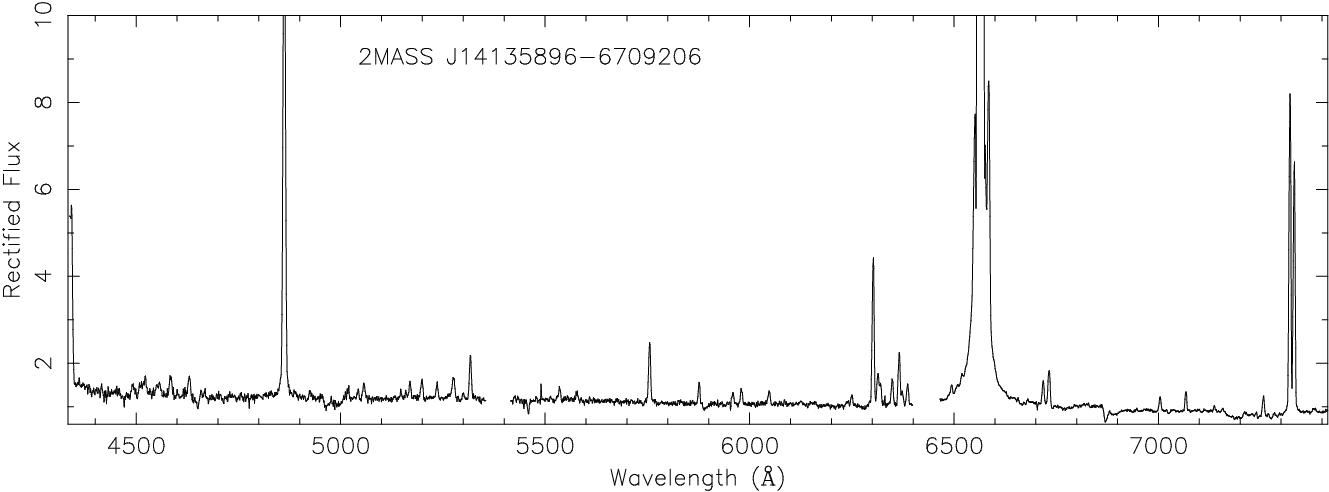}
\includegraphics[scale=0.55,angle=270]{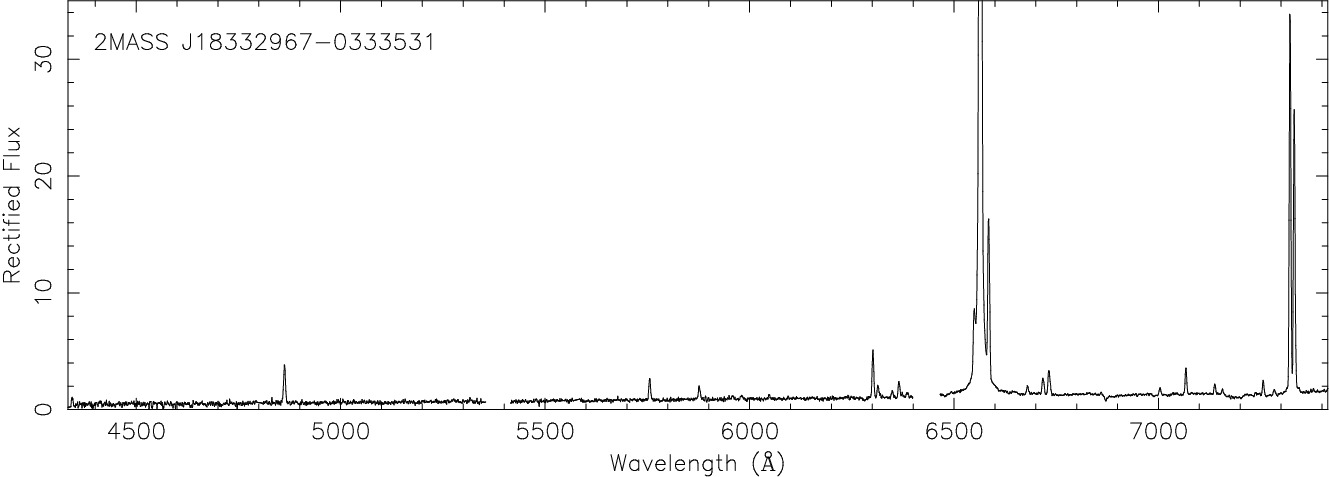}
   \end{center}
   \caption{SALT RSS spectra of the B[e] star 2MASS J18500902$-$0515041 and the possible Be stars 2MASS J14135896$-$6709206 and 2MASS J18332967$-$0333531.}
   \label{fig:be2}
\end{figure*}

The most numerous in Tab. \ref{tab:other} are six B[e] stars whose spectra are displayed in Figures \ref{fig:be1} and \ref{fig:be2}. Their spectra are similar to two B[e] stars found by Corradi et al. (2010a) and one by MMU13. They are classified according to the criteria given by Lamers et al. (1998), namely the presence of strong Balmer emission lines, low excitation permitted emission lines (e.g. Fe~II), forbidden emission lines of [Fe~II] and [O~I], and a strong NIR excess due to hot dust. The last property is evident in Fig. \ref{fig:2mass} where the B[e] stars cluster around $J-H\sim1.6$ and $H-K_s\sim1.7$, except for 2MASS J18285064$-$2432017 which has redder colours. All six show fluorescent [Ni~II] emission lines $\lambda\lambda$6667, 7378 and 7412 \AA\ (Lucy 1995; Corradi et al. 2010a). The relatively high number of B[e] stars found can be explained by the higher priority we gave to dusty D-type candidates in the SALT queue. This was done in order to improve the chances of finding D-type symbiotics, which have proven difficult to find (Corradi et al. 2008, 2010a; MMU13), but this was complicated by several object types sharing H$\alpha$ emission and hot dust. 

A search of SIMBAD and Vizier of the six B[e] stars showed that some were misclassified as PNe. Stenholm \& Acker (1987) commented that only H$\alpha$ was visible in their spectra for 2MASS J18285064$-$2432017 (H~2-47) and 2MASS J18500902$-$0515041 (THA 14-39, The 1962). The B[e] star 2MASS J16092019$-$5536100 was originally catalogued as a possible new PN in the IRAS catalogue (IRAS 16053-5528; PM 1-106, Preite-Martinez et al. 1988). Later, it appeared in Su{\'a}rez et al. (2006) catalogue where it is classified as a low excitation class PN. Mottram et al. (2011) remarked that 2MASS J16141537$-$5146036 (IRAS 16103-5138) is more likely a young or old star, rather than an HII region or PN. 

Finally, 2MASS J14135896$-$6709206, listed as the H$\alpha$ emission line star SS 255 (Stephenson \& Sanduleak 1977a), strongly resembles the B[e] stars, however the 2MASS colours do not show evidence for hot dust and the [Ni~II] emission lines are absent. Similar objects were studied by Pereira et al. (2003, 2008) and their classification is particularly difficult. We suggest a Be star (Porter \& Rivinius 2003) classification for 2MASS J14135896$-$6709206, based on the similar appearance to SS73 24 (Pereira et al. 2003). Given the strong similarity between the spectra of 2MASS J14135896$-$6709206 and 2MASS J18332967$-$0333531, we suggest that 2MASS J18332967$-$0333531 is a reddened Be star. Other possible classifications for these two objects include a B[e] star with no hot dust (e.g. Graus, Lamb \& Oey 2012) or a pre-PN (Pereira et al. 2008), but it is beyond the scope of this work to speculate further on this point. 

\subsection[]{Wolf-Rayet stars}
\label{sec:wr}
\begin{figure*}
   \begin{center}
\includegraphics[scale=0.55,angle=270]{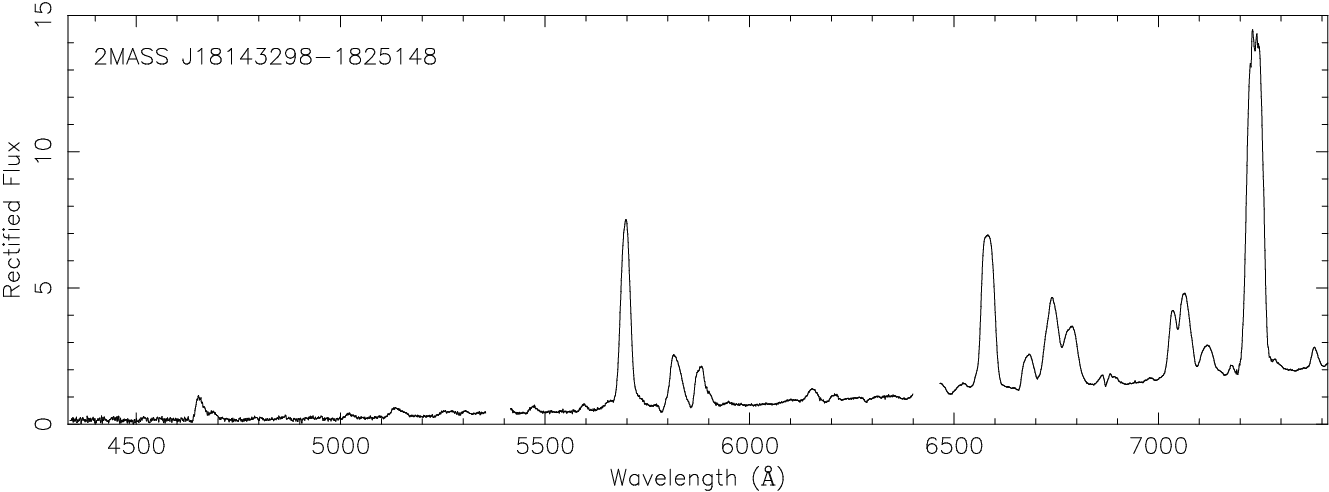}
\includegraphics[scale=0.55,angle=270]{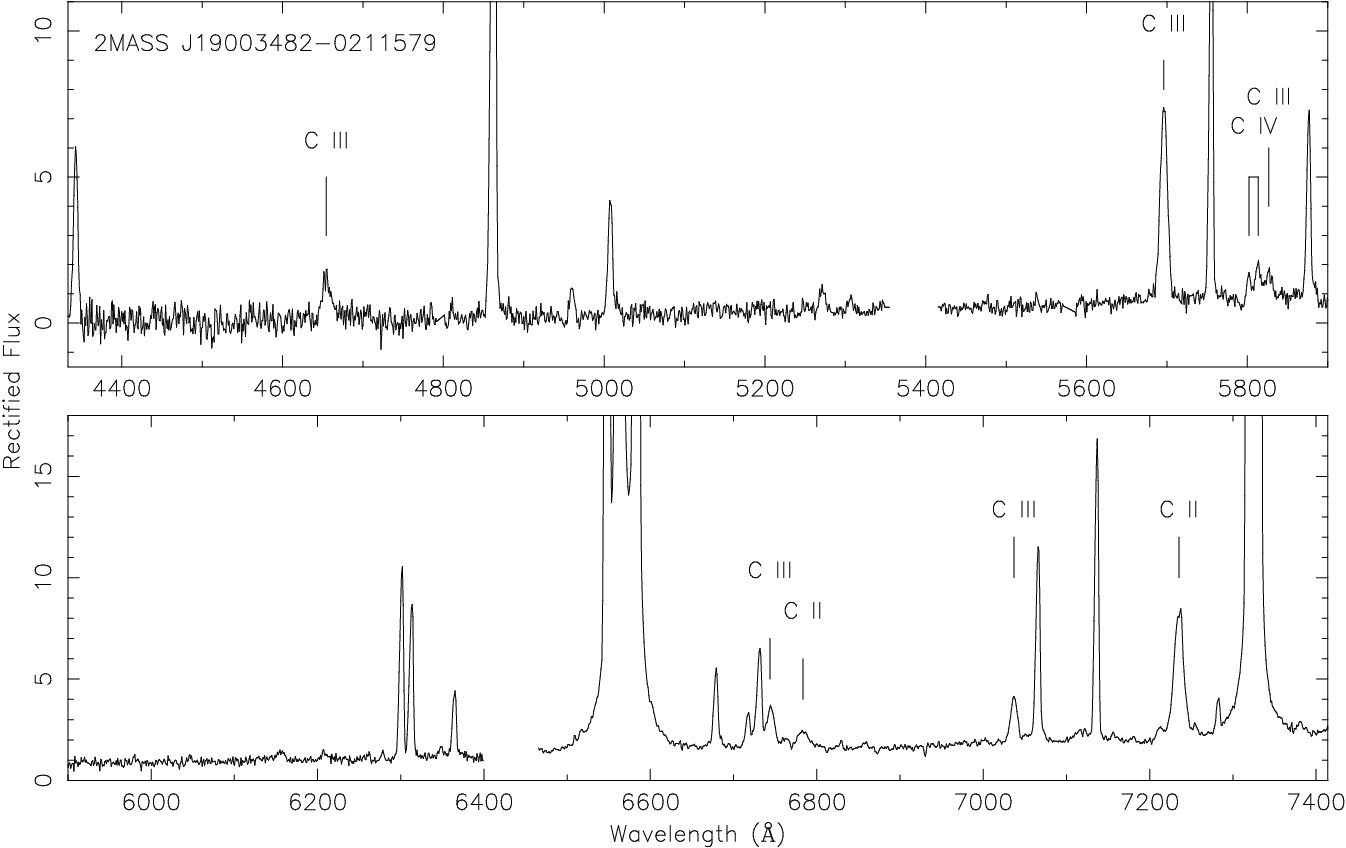}
   \end{center}
   \caption{SALT RSS spectrum of the massive WC9 star 2MASS J18143298$-$1825148 and the PN K3-18 (2MASS J19003482$-$0211579) revealing its [WC9] central star.}
   \label{fig:wr}
\end{figure*}

The SALT RSS spectrum of 2MASS J18143298$-$1825148 in Fig. \ref{fig:wr}, originally catalogued as emission line star THA 34-30 (The 1966), is typical of a reddened WC9 star (Crowther et al. 1998).\footnote{The bright C~II $\lambda$7235 emission line is a scaled replacement from the short exposure where the line core was unsaturated.} We measured the equivalent widths of O~V $\lambda$5590, C~III $\lambda$5696 and C~IV $\lambda\lambda$5801--5812 to find log (C~IV/C~III)=$-$0.50 and log (C~III/O~V)=1.62, values that represent a clear WC9 classification in figure 6 of Crowther et al. (1998). While there are many Galactic WC9 stars known (e.g. van der Hucht 2001), it is interesting to note that dusty WC9 stars are more difficult to discover in NIR selected samples (Shara et al. 2009, 2012; Mauerhan et al. 2009, 2011). The Shara et al. survey field containing 2MASS J18143298$-$1825148 is missing several narrowband images, and this has not been searched for new WR stars (Shara, private communication).

The object 2MASS J19003482$-$0211579 was first discovered as the PN K3-18 (Kohoutek 1965). Stenholm \& Acker (1987) suggested the nebula was not a PN, noting abnormally strong [O~II] emission lines, while Giammanco et al. (2011) previously suspected K3-18 to be a symbiotic star based on its red 2MASS colours. In conjunction with narrow nebular emission lines of the PN, the SALT RSS spectrum in Fig. \ref{fig:wr} clearly shows broad stellar emission lines from carbon species, typical of low mass WR central stars of PNe (Crowther et al. 1998; G\'orny et al. 2004, 2009; Acker \& Neiner 2003; DePew et al. 2011; Miszalski et al. 2011c; MMU13). To classify the central star we again use the scheme of Crowther et al. (1998). In the C~III $\lambda$5696 line we measure a FWHM=$10.2\pm0.1$ \AA\ and an equivalent width $W_\lambda=-75\pm5$ \AA. The C~IV $\lambda$5801,5812 lines are resolved with equivalent widths of approximately $-$7 and $-$13 \AA, respectively. If we take the log $W_\lambda$ ratio of both C~IV lines to C~III $\lambda$5696 we get $-0.55$, which falls inside the $-0.7$ to $-0.3$ range for the WC9 subtype (Crowther et al. 1998). As this is the primary classification criterion, we adopt a [WC9] classification for the central star, where the square brackets distinguish WR central stars of PNe apart from massive WR stars.

\subsection[]{Planetary Nebulae}
\begin{figure*}
   \begin{center}
\includegraphics[scale=0.6,angle=270]{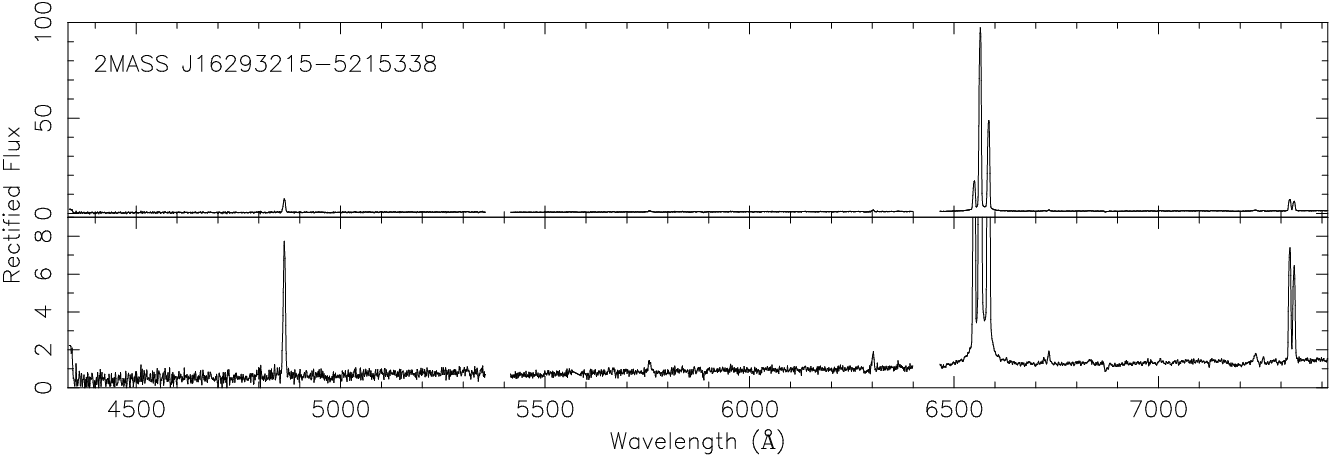}
\includegraphics[scale=0.6,angle=270]{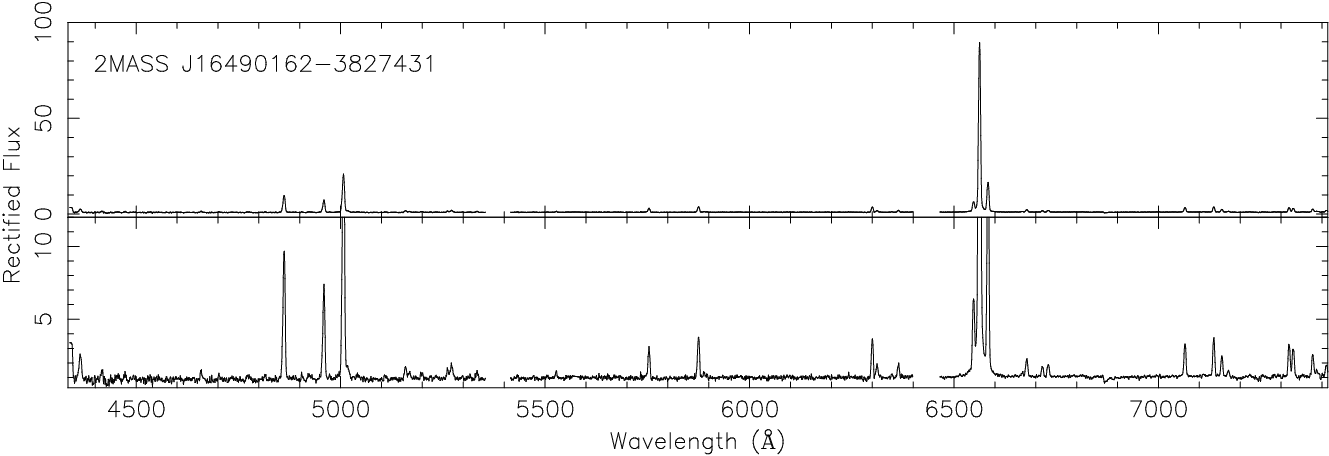}
\includegraphics[scale=0.6,angle=270]{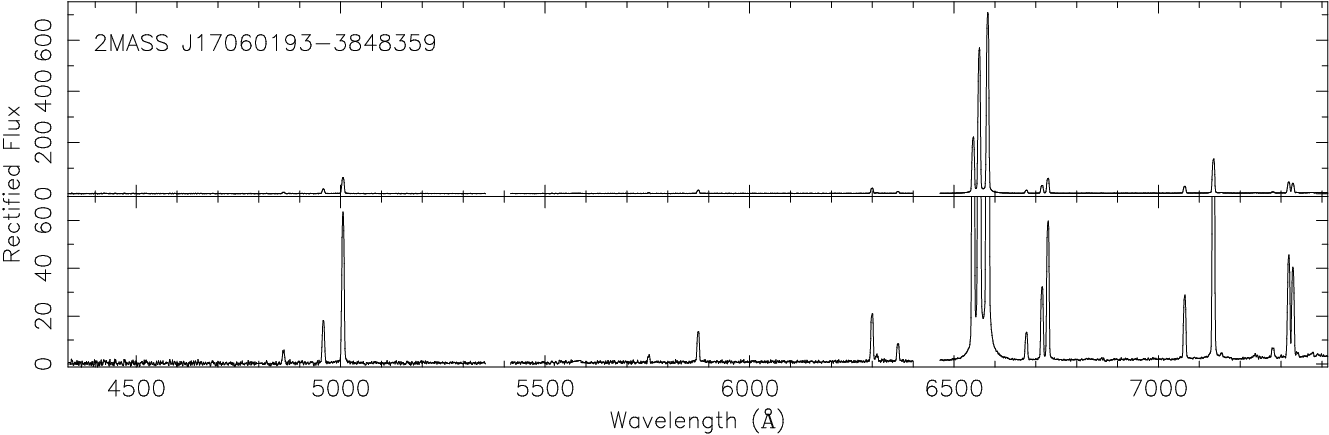}
\includegraphics[scale=0.6,angle=270]{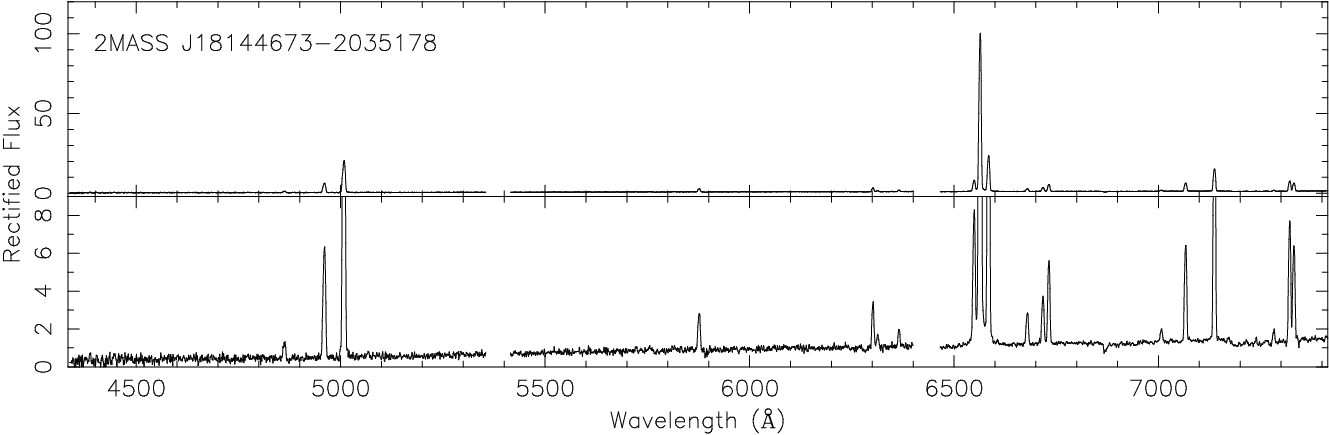}
\includegraphics[scale=0.6,angle=270]{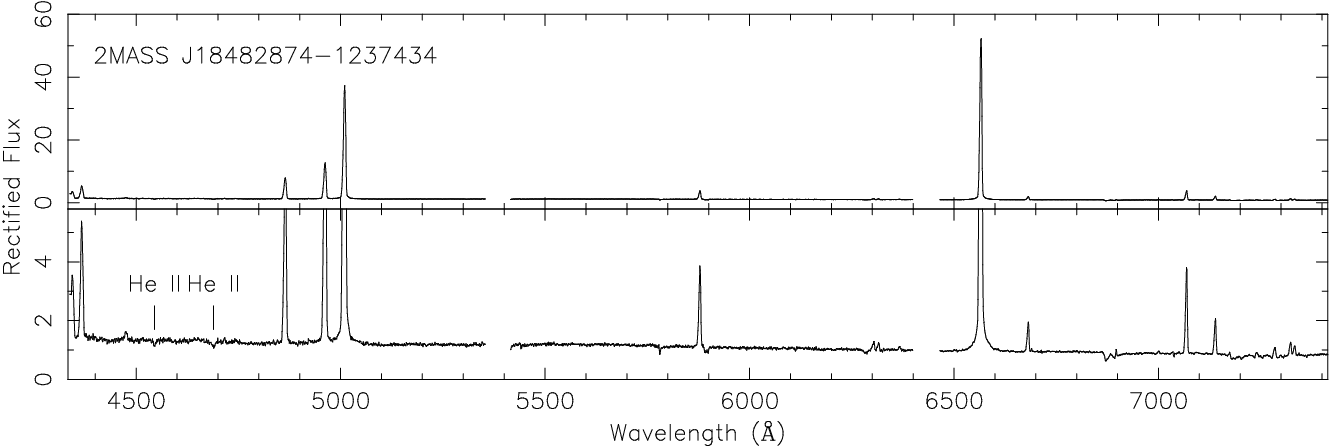}
   \end{center}
   \caption{SALT RSS spectra of PNe or PN-like objects.}
   \label{fig:pn1}
\end{figure*}

Reddened PNe may also be found near D-type symbiotic stars in Fig. \ref{fig:2mass}. Figure \ref{fig:pn1} displays SALT RSS spectra of objects suspected to be reddened PNe in our sample. It can be difficult, however, to distinguish them from reddened compact or ultra-compact HII regions (e.g. Cohen et al. 2011). A useful diagnostic in this respect is the MIR/radio flux ratio (Cohen et al. 2011). Low values of the ratio are common amongst bona-fide PNe, whereas high values are more common amongst HII regions. Taking the \emph{AKARI} 9 $\mu$m (Ishihara et al. 2010) and NVSS 1.4 GHz (Condon et al. 1998) fluxes for 2MASS J17060193$-$3848359, we derive a ratio of 6.32. This low ratio, combined with the relatively strong ([N~II]$\lambda$6548+[N~II]$\lambda$6583)/H$\alpha$ ratio of 1.64, suggests a firm PN classification, contrary to the candidate YSO classification given by Mottram et al. (2007). The situation is less clear for 2MASS J18144673$-$2035178 which has a relatively high ratio of 18, calculated using the GLIMPSE band 4 (8.0$\mu$m) flux (Benjamin et al. 2003; Churchwell et al. 2009) instead of an \emph{AKARI} 9 $\mu$m flux. The [O~III] emission and the log (H$\alpha$/[N~II]) and log (H$\alpha$/[S~II]) line ratios of 0.51 and 1.18, respectively, suggest a PN classification (e.g. Tajitsu et al. 1999).

Other objects do not have the required data available to calculate their MIR/radio ratios. The emission line pattern in 2MASS J16490162$-$3827431 appears to be typical of PNe, so we consider it to be a PN. While the absence of [O~III] in 2MASS J16293215$-$5215338 is suspicious, it otherwise appears PN-like, perhaps a PN with a cool ($T_\mathrm{eff}\sim30$ kK) central star, explaining our possible PN classification. A clear PN classification is made for 2MASS J18482874$-$1237434 whose hot central star is detected in the SALT RSS spectrum. Both He~II $\lambda$4540 and $\lambda$4686 are detected in absorption, together with a blue continuum, and He~II $\lambda$5412 may also be present on the edge of the chip gap. The emission line pattern is somewhat unusual for a PN with the intensity of [O~III] $\lambda$4363 exceeding H$\gamma$, indicating a high density environment. This places 2MASS J18482874$-$1237434 amongst a growing group of PNe with a high density core (HDC) that may be symptomatic of some kind of dusty, wide binary interaction with a main-sequence companion (see e.g. Liebert et al. 1989, 2013; Miszalski et al. 2011a,b; MMU13).

\subsection[]{Young stellar objects}
\begin{figure*}
   \begin{center}
\includegraphics[scale=0.55,angle=270]{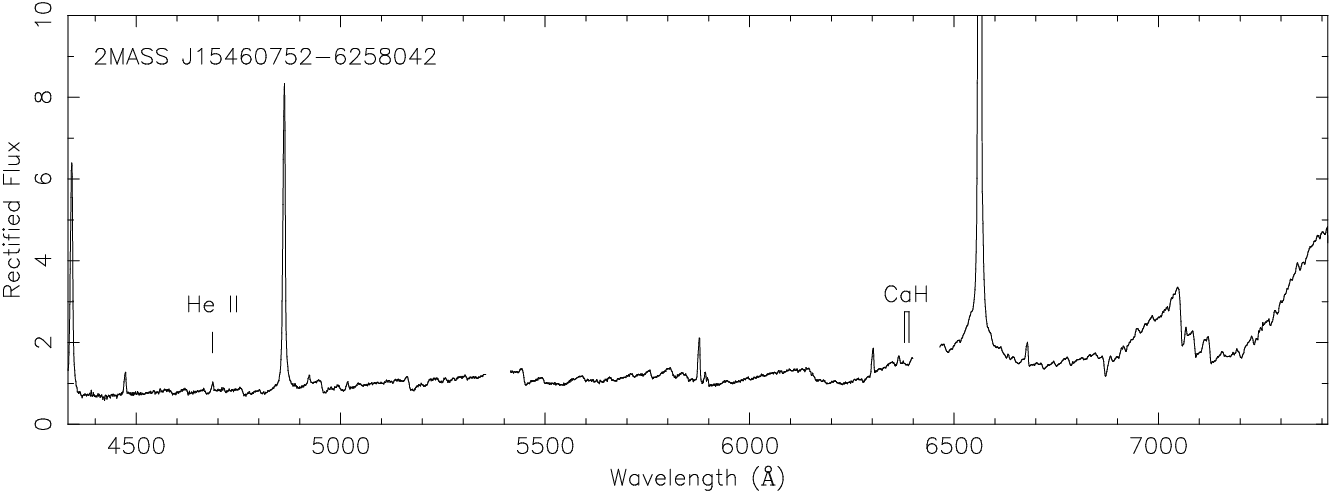}
\includegraphics[scale=0.55,angle=270]{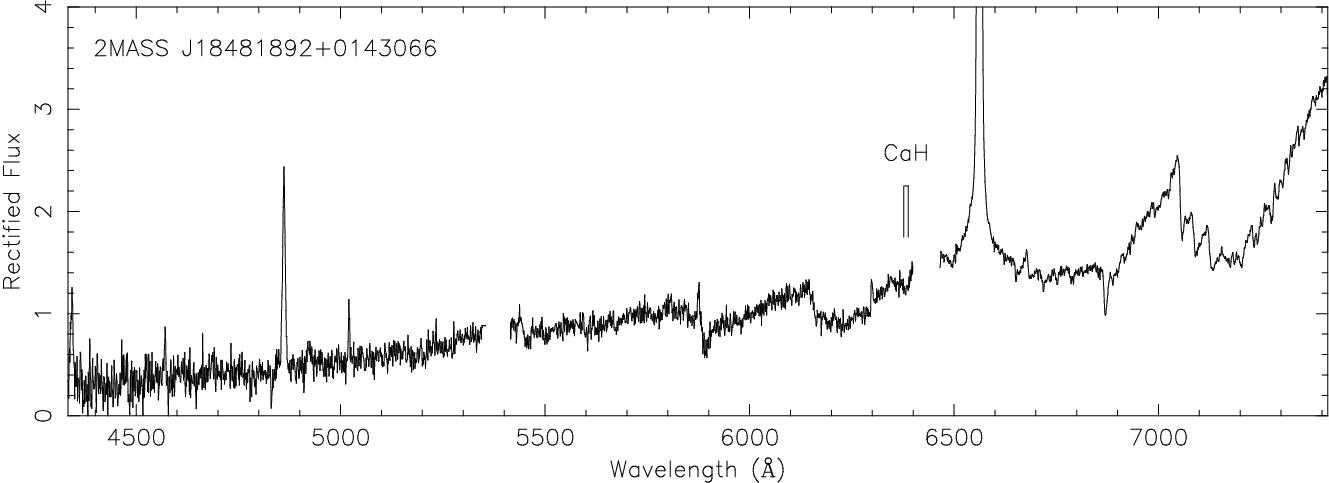}
\includegraphics[scale=0.55,angle=270]{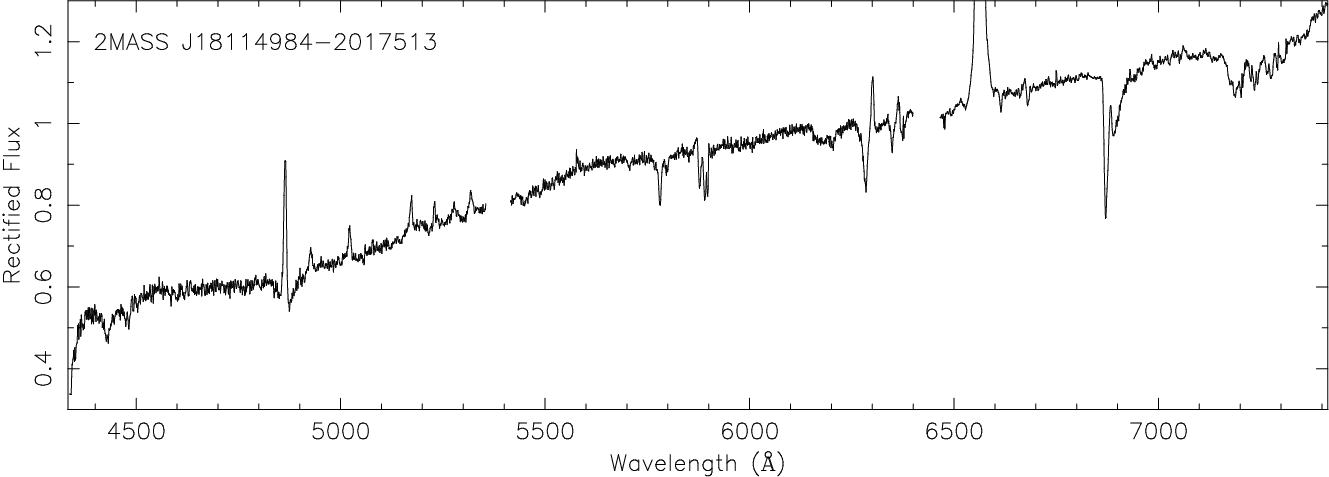}
   \end{center}
   \caption{SALT RSS spectra of young stellar objects.}
   \label{fig:o1}
\end{figure*}

Some emission line stars can prove challenging to classify in symbiotic star surveys. Young stellar objects, in particular those few examples showing He~II $\lambda$4686 emission (e.g. Costa et al. 1999), can appear very similar to symbiotic star spectra. Figure \ref{fig:o1} shows three examples of young stellar objects identified, two T Tauri stars (2MASS J15460752$-$6258042 and 2MASS J18481892$+$0143066) and one Herbig Ae/Be star (2MASS J18114984$-$2017513). Both of the T Tauri stars have clearly identifiable CaH $\lambda\lambda$ 6382, 6389 \AA\ absorption bands that are found in T Tauri stars and never in red giants (e.g. Cohen \& Kuhi 1979). This is an additional useful diagnostic to the NIR colours that also indicate a reddened luminosity class V-IV star, rather than a giant star (see Fig. \ref{fig:2mass}). While the emission line spectrum of 2MASS J15460752$-$6258042 does show weak HeII 4686 emission, no other high exctitation lines are present. The presence of Na~I D lines in emission is common in T Tauri stars, but relatively rare in symbiotic stars (an exception is RX Pup, Miko{\l}ajewska et al. 1999). We have also identified 2MASS J18114984$-$2017513 as a Herbig Ae/Be star using the online atlas of R.O. Gray.\footnote{http://ned.ipac.caltech.edu/level5/Gray/Gray16.html} This object demonstrates more reddened NIR colours, consistent with it being embedded in more dust. All these factors, combined with those discussed by Corradi et al. (2010a), serve as useful diagnostics to distinguish some young stellar objects from symbiotic stars.

\begin{thebibliography}{}

\bibitem[Acker \& Neiner(2003)]{2003A&A...403..659A} Acker, A., \& Neiner, C.\ 2003, A\&A, 403, 659 
\bibitem[Allen(1984)]{1984PASAu...5..369A} Allen, D.~A.\ 1984, PASA, 5, 369 

\bibitem[Andrillat \& Houziaux(1994)]{1994MNRAS.271..875A} Andrillat, Y., \& Houziaux, L.\ 1994, MNRAS, 271, 875 
\bibitem[\protect\citeauthoryear{Barnbaum, Stone, \& Keenan}{1996}]{1996ApJS..105..419B} Barnbaum C., Stone R.~P.~S., Keenan P.~C., 1996, ApJS, 105, 419 
      
\bibitem[Belczy{\'n}ski et al.(2000)]{2000A&AS..146..407B} Belczy{\'n}ski, K., Miko{\l}ajewska, J., Munari, U., Ivison, R.~J., \& Friedjung, M.\ 2000, A\&AS, 146, 407 

\bibitem[Benjamin et al.(2003)]{2003PASP..115..953B} Benjamin, R.~A., Churchwell, E., Babler, B.~L., et al.\ 2003, PASP, 115, 953 

\bibitem[Bessell \& Brett(1988)]{1988PASP..100.1134B} Bessell, M.~S., \& Brett, J.~M.\ 1988, PASP, 100, 1134 


   

\bibitem[Buckley et al.(2006)]{2006SPIE.6267E..32B} Buckley, D.~A.~H., Swart, G.~P., \& Meiring, J.~G.\ 2006, SPIE, 6267, 32  

\bibitem[Burgh et al.(2003)]{2003SPIE.4841.1463B} Burgh, E.~B., Nordsieck, K.~H., Kobulnicky, H.~A., et al.\ 2003, SPIE, 4841, 1463 






\bibitem[Cardelli et al.(1989)]{1989ApJ...345..245C} Cardelli, J.~A., Clayton, G.~C., \& Mathis, J.~S.\ 1989, ApJ, 345, 245 

\bibitem[Carpenter(2001)]{2001AJ....121.2851C} Carpenter, J.~M.\ 2001, AJ, 121, 2851 


   

\bibitem[Churchwell et al.(2009)]{2009PASP..121..213C} Churchwell, E., Babler, B.~L., Meade, M.~R., et al.\ 2009, PASP, 121, 213 

\bibitem[Cohen \& Kuhi(1979)]{1979ApJS...41..743C} Cohen, M., \& Kuhi, L.~V.\ 1979, ApJS, 41, 743 


   

\bibitem[Cohen et al.(2011)]{2011MNRAS.413..514C} Cohen, M., Parker, Q.~A., Green, A.~J., et al.\ 2011, MNRAS, 413, 514 


\bibitem[Condon et al.(1998)]{1998AJ....115.1693C} Condon, J.~J., Cotton, W.~D., Greisen, E.~W., et al.\ 1998, AJ, 115, 1693 

   
\bibitem[Corradi et al.(2008)]{2008A&A...480..409C} Corradi, R.~L.~M., Rodr{\'{\i}}guez-Flores, E.~R., Mampaso, A., et al.\ 2008, A\&A, 480, 409 
\bibitem[Corradi \& Giammanco(2010)]{2010A&A...520A..99C} Corradi, R.~L.~M., \& Giammanco, C.\ 2010, A\&A, 520, A99 
\bibitem[Corradi et al.(2010)]{2010A&A...509A..41C} Corradi, R.~L.~M., Valentini, M., Munari, U., et al.\ 2010a, A\&A, 509, A41 
\bibitem[Corradi et al.(2010)]{2010A&A...509L...9C} Corradi, R.~L.~M., Munari, U., Greimel, R., et al.\ 2010b, A\&A, 509, L9 

\bibitem[Corradi et al.(2011)]{2011A&A...529A..56C} Corradi, R.~L.~M., Sabin, L., Munari, U., et al.\ 2011, A\&A, 529, A56 
\bibitem[Corradi(2012)]{2012BaltA..21...32C} Corradi, R.~L.~M.\ 2012, Baltic Astronomy, 21, 32 


\bibitem[Costa et al.(1999)]{1999MNRAS.307L..23C} Costa, V.~M., Gameiro, J.~F., \& Lago, M.~T.~V.~T.\ 1999, MNRAS, 307, L23 



   
\bibitem[Crawford et al.(2010)]{2010SPIE.7737E..54C} Crawford, S.~M., Still, M., Schellart, P., et al.\ 2010, Proc. SPIE, 7737E, 54 

\bibitem[Crowther et al.(1998)]{1998MNRAS.296..367C} Crowther, P.~A., De Marco, O., \& Barlow, M.~J.\ 1998, MNRAS, 296, 367 




\bibitem[Dilday et al.(2012)]{2012Sci...337..942D} Dilday, B., Howell, D.~A., Cenko, S.~B., et al.\ 2012, Science, 337, 942 

\bibitem[Di Stefano(2010)]{2010ApJ...719..474D} Di Stefano, R.\ 2010, ApJ, 719, 474 

\bibitem[Depew et al.(2011)]{2011MNRAS.414.2812D} Depew, K., Parker, Q.~A., Miszalski, B., et al.\ 2011, MNRAS, 414, 2812 


\bibitem[Drew et al.(2005)]{2005MNRAS.362..753D} Drew, J.~E., Greimel, R., Irwin, M.~J., et al.\ 2005, MNRAS, 362, 753 




\bibitem[\protect\citeauthoryear{Escudero \& Costa}{2001}]{2001A&A...380..300E} Escudero A.~V., Costa R.~D.~D., 2001, A\&A, 380, 300 


\bibitem[Giammanco et al.(2011)]{2011A&A...525A..58G} Giammanco, C., Sale, S.~E., Corradi, R.~L.~M., et al.\ 2011, A\&A, 525, A58 



\bibitem[G{\'o}rny et al.(2004)]{2004A&A...427..231G} G{\'o}rny, S.~K., Stasi{\'n}ska, G., Escudero, A.~V., \& Costa, R.~D.~D.\ 2004, A\&A, 427, 231 



\bibitem[G{\'o}rny et al.(2009)]{2009A&A...500.1089G} G{\'o}rny, S.~K., Chiappini, C., Stasi{\'n}ska, G., \& Cuisinier, F.\ 2009, A\&A, 500, 1089 

\bibitem[Graus et al.(2012)]{2012ApJ...759...10G} Graus, A.~S., Lamb, J.~B., \& Oey, M.~S.\ 2012, ApJ, 759, 10 





   

\bibitem[Gromadzki et al.(2009)]{2009AcA....59..169G} Gromadzki, M., Miko{\l}ajewska, J., Whitelock, P., \& Marang, F.\ 2009, Acta Astron., 59, 169 


\bibitem[Gromadzki et al.(2013)]{2013arXiv1312.6063G} Gromadzki, M., Miko{\l}ajewska, J., \& Soszy{\'n}ski, I.\ 2013, Acta Astron., 63, 405


\bibitem[Hamuy et al.(1994)]{1994PASP..106..566H} Hamuy, M., Suntzeff, N.~B., Heathcote, S.~R., et al.\ 1994, PASP, 106, 566 


\bibitem[Ishihara et al.(2010)]{2010A&A...514A...1I} Ishihara, D., Onaka, T., Kataza, H., et al.\ 2010, A\&A, 514, A1 

\bibitem[Jordan et al.(1996)]{1996A&A...312..897J} Jordan, S., Schmutz, W., Wolff, B., Werner, K., \& Muerset, U.\ 1996, A\&A, 312, 897 




\bibitem[Kato et al.(2013)]{2013ApJ...763....5K} Kato, M., Hachisu, I., \& Miko{\l}ajewska, J.\ 2013, ApJ, 763, 5 

   


\bibitem[Keenan(1993)]{1993PASP..105..905K} Keenan, P.~C.\ 1993, PASP, 105, 905 

   

\bibitem[Kenyon(1986)]{1986syst.book.....K} Kenyon, S.~J.\ 1986, Cambridge and New York, Cambridge University Press, 1986, 295 p.

\bibitem[Kenyon \& Fernandez-Castro(1987)]{1987AJ.....93..938K} Kenyon, S.~J., \& Fernandez-Castro, T.\ 1987, AJ, 93, 938 


\bibitem[Kenyon et al.(1993)]{1993ApJ...407L..81K} Kenyon, S.~J., Livio, M., Miko{\l}ajewska, J., \& Tout, C.~A.\ 1993, ApJL, 407, L81 

   


\bibitem[Kobulnicky et al.(2003)]{2003SPIE.4841.1634K} Kobulnicky, H.~A., Nordsieck, K.~H., Burgh, E.~B., et al.\ 2003, SPIE, 4841, 1634 

\bibitem[Kohoutek(1965)]{1965BAICz..16..221K} Kohoutek, L.\ 1965, Bulletin of the Astronomical Institutes of Czechoslovakia, 16, 221 
   


\bibitem[Kohoutek(1994)]{1994AN....315..235K} Kohoutek, L.\ 1994, Astronomische Nachrichten, 315, 235 
\bibitem[Kohoutek \& Wehmeyer(2003)]{2003AN....324..437K} Kohoutek, L., \& Wehmeyer, R.\ 2003, Astronomische Nachrichten, 324, 437 



   
\bibitem[Lamers et al.(1998)]{1998A&A...340..117L} Lamers, H.~J.~G.~L.~M., Zickgraf, F.-J., de Winter, D., Houziaux, L., \& Zorec, J.\ 1998, A\&A, 340, 117 

\bibitem[Liebert et al.(1989)]{1989ApJ...346..251L} Liebert, J., Green, R., Bond, H.~E., et al.\ 1989, ApJ, 346, 251 

\bibitem[Liebert et al.(2013)]{2013ApJ...769...32L} Liebert, J., Bond, H.~E., Dufour, P., et al.\ 2013, ApJ, 769, 32 


\bibitem[\protect\citeauthoryear{Lucy}{1995}]{1995A&A...294..555L} Lucy L.~B., 1995, A\&A, 294, 555 



\bibitem[Luna et al.(2013)]{2013A&A...559A...6L} Luna, G.~J.~M., Sokoloski, J.~L., Mukai, K., \& Nelson, T.\ 2013, A\&A, 559, A6 


\bibitem[Magrini et al.(2003)]{2003ASPC..303..539M} Magrini, L., Corradi, R.~L.~M., \& Munari, U.\ 2003, Symbiotic Stars Probing Stellar Evolution, 303, 539 
   
\bibitem[\protect\citeauthoryear{Mauerhan, van Dyk, \& Morris}{2009}]{2009PASP..121..591M} Mauerhan J.~C., van Dyk S.~D., Morris P.~W., 2009, PASP, 121, 591 

\bibitem[Mauerhan et al.(2011)]{2011AJ....142...40M} Mauerhan, J.~C., Van Dyk, S.~D., \& Morris, P.~W.\ 2011, AJ, 142, 40 

   


\bibitem[McLaughlin(1953)]{1953ApJ...118...27M} McLaughlin, D.~B.\ 1953, ApJ, 118, 27 


\bibitem[Mikolajewska et al.(1995)]{1995AJ....109.1289M} Miko{\l}ajewska, J., Kenyon, S.~J., Miko{\l}ajewski, M., Garcia, M.~R., \& Polidan, R.~S.\ 1995, AJ, 109, 1289 





\bibitem[Mikolajewska et al.(1997)]{1997A&A...327..191M} Miko{\l}ajewska, J., Acker, A., \& Stenholm, B.\ 1997, A\&A, 327, 191 



\bibitem[Mikolajewska et al.(1999)]{1999MNRAS.305..190M} Miko{\l}ajewska, J., Brandi, E., Hack, W., et al.\ 1999, MNRAS, 305, 190 


\bibitem[Miko{\l}ajewska(2003)]{2003ASPC..303....9M} Miko{\l}ajewska, J.\ 2003, Symbiotic Stars Probing Stellar Evolution, 303, 9 




\bibitem[Miko{\l}ajewska(2004)]{2004RMxAC..20...33M} Miko{\l}ajewska, J.\ 2004, Revista Mexicana de Astronomia y Astrofisica Conference Series, 20, 33 





\bibitem[Miko{\l}ajewska(2012)]{2012BaltA..21....5M} Miko{\l}ajewska, J.\ 2012, Baltic Astronomy, 21, 5 

\bibitem[Miszalski et al.(2008)]{2008MNRAS.384..525M} Miszalski, B., Parker, Q.~A., Acker, A., et al.\ 2008, MNRAS, 384, 525 

\bibitem[Miszalski et al.(2011)]{2011A&A...528A..39M} Miszalski, B., Miko{\l}ajewska, J., K{\"o}ppen, J., et al.\ 2011a, A\&A, 528, A39 

\bibitem[Miszalski et al.(2011)]{2011apn5.confP.109M} Miszalski, B., Acker, A., Parker, Q.~A., et al.\ 2011b, Asymmetric Planetary Nebulae 5 Conference, 109P 

\bibitem[Miszalski et al.(2011)]{2011A&A...529A..77M} Miszalski, B., Napiwotzki, R., Cioni, M.-R.~L., \& Nie, J.\ 2011c, A\&A, 529, A77 


\bibitem[Miszalski et al.(2013)]{2013MNRAS.432.3186M} Miszalski, B., Miko{\l}ajewska, J., \& Udalski, A.\ 2013, MNRAS, 432, 3186 (MMU13)


\bibitem[Mottram et al.(2007)]{2007A&A...476.1019M} Mottram, J.~C., Hoare, M.~G., Lumsden, S.~L., et al.\ 2007, A\&A, 476, 1019 




\bibitem[Mottram et al.(2011)]{2011A&A...525A.149M} Mottram, J.~C., Hoare, M.~G., Urquhart, J.~S., et al.\ 2011, A\&A, 525, A149 


\bibitem[Muerset et al.(1997)]{1997A&A...319..201M} M{\"u}rset, U., Wolff, B., \& Jordan, S.\ 1997, A\&A, 319, 201 

\bibitem[Munari \& Renzini(1992)]{1992ApJ...397L..87M} Munari, U., \& Renzini, A.\ 1992, ApJL, 397, L87 

\bibitem[\protect\citeauthoryear{Nielsen et al.}{2013}]{2013arXiv1310.2170N} Nielsen M.~T.~B., Nelemans G., Voss R., Toonen S., 2013, A\&A, in press, arXiv:1310.2170 

   

\bibitem[O'Donoghue et al.(2006)]{2006MNRAS.372..151O} O'Donoghue, D., Buckley, D.~A.~H., Balona, L.~A., et al.\ 2006, MNRAS, 372, 151 


\bibitem[Orio et al.(2007)]{2007ApJ...661.1105O} Orio, M., Zezas, A., Munari, U., Siviero, A., \& Tepedelenlioglu, E.\ 2007, ApJ, 661, 1105 

\bibitem[Parker et al.(2005)]{2005MNRAS.362..689P} Parker, Q.~A., Phillipps, S., Pierce, M.~J., et al.\ 2005, MNRAS, 362, 689 

\bibitem[Pereira et al.(2003)]{2003A&A...397..927P} Pereira, C.~B., Franco, C.~S., \& de Ara{\'u}jo, F.~X.\ 2003, A\&A, 397, 927 

   
\bibitem[Pereira et al.(2008)]{2008A&A...477..877P} Pereira, C.~B., Marcolino, W.~L.~F., Machado, M., \& de Ara{\'u}jo, F.~X.\ 2008, A\&A, 477, 877 




\bibitem[Pierce (2005)]{2005Pierce} Pierce, M.J. 2005, PhD thesis, University of Bristol
   

\bibitem[Porter \& Rivinius(2003)]{2003PASP..115.1153P} Porter, J.~M., \& Rivinius, T.\ 2003, PASP, 115, 1153 

\bibitem[Preite-Martinez(1988)]{1988A&AS...76..317P} Preite-Martinez, A.\ 1988, A\&AS, 76, 317 


\bibitem[Schlafly \& Finkbeiner(2011)]{2011ApJ...737..103S} Schlafly, E.~F., \& Finkbeiner, D.~P.\ 2011, ApJ, 737, 103 

\bibitem[Schmeja \& Kimeswenger(2001)]{2001A&A...377L..18S} Schmeja, S., \& Kimeswenger, S.\ 2001, A\&A, 377, L18 

\bibitem[Schmid(1989)]{1989A&A...211L..31S} Schmid, H.~M.\ 1989, A\&A, 211, L31 
   
\bibitem[\protect\citeauthoryear{Shara et al.}{2009}]{2009AJ....138..402S} Shara M.~M., et al., 2009, AJ, 138, 402 
   
\bibitem[\protect\citeauthoryear{Shara et al.}{2012}]{2012AJ....143..149S} Shara M.~M., Faherty J.~K., Zurek D., Moffat A.~F.~J., Gerke J., Doyon R., Artigau E., Drissen L., 2012, AJ, 143, 149 

\bibitem[Skiff]{2013Skiffviz} Skiff, B.~A.\ 2013,  Catalogue of Stellar Spectral Classifications, VizieR Online Data Catalog, 1, 2023, \verb|http://cdsarc.u-strasbg.fr/viz-bin/Cat?B/mk|

\bibitem[Skrutskie et al.(2006)]{2006AJ....131.1163S} Skrutskie, M.~F., Cutri, R.~M., Stiening, R., et al.\ 2006, AJ, 131, 1163 
   

\bibitem[\protect\citeauthoryear{Soszynski et al.}{2013}]{2013arXiv1304.2787S} Soszy{\'n}ski I., et al., 2013, Acta Astron., 63, 21



\bibitem[Stenholm \& Acker(1987)]{1987A&AS...68...51S} Stenholm, B., \& Acker, A.\ 1987, A\&AS, 68, 51 

\bibitem[Stephenson \& Sanduleak(1977)]{1977ApJS...33..459S} Stephenson, C.~B., \& Sanduleak, N.\ 1977a, ApJS, 33, 459 

\bibitem[Stephenson \& Sanduleak(1977)]{1977PW&SO...2...71S} Stephenson, C.~B., \& Sanduleak, N.\ 1977b, Publications of the Warner \& Swasey Observatory, 2, 71 
   
\bibitem[Su{\'a}rez et al.(2006)]{2006A&A...458..173S} Su{\'a}rez, O., Garc{\'{\i}}a-Lario, P., Manchado, A., et al.\ 2006, A\&A, 458, 173 
   

\bibitem[Tajitsu et al.(1999)]{1999PASP..111.1157T} Tajitsu, A., Tamura, S., Yadoumaru, Y., Weinberger, R., K{\"o}ppen, J.\ 1999, PASP, 111, 1157 

\bibitem[\protect\citeauthoryear{Taylor}{2006}]{2006ASPC..351..666T} Taylor M.~B., 2006, ASPC, 351, 666 
\bibitem[\protect\citeauthoryear{Taylor}{2011}]{2011ascl.soft05001T} Taylor M., 2011, ascl.soft, 5001 


\bibitem[Terzan \& Gosset(1991)]{1991A&AS...90..451T} Terzan, A., \& Gosset, E.\ 1991, A\&AS, 90, 451 



\bibitem[The(1962)]{1962CoBos..14....0T} The, P.~S.\ 1962, Contributions from the Bosscha Observervatory, 14, 0 
\bibitem[The(1966)]{1966CoBos..34....1T} The, P.-S.\ 1966, Contributions from the Bosscha Observervatory, 34, 1 


\bibitem[van der Hucht(2001)]{2001NewAR..45..135V} van der Hucht, K.~A.\ 2001, NewAR, 45, 135 
   
\bibitem[van Dokkum(2001)]{2001PASP..113.1420V} van Dokkum, P.~G.\ 2001, PASP, 113, 1420 

\bibitem[Whitelock(1987)]{1987PASP...99..573W} Whitelock, P.~A.\ 1987, PASP, 99, 573 




   
\bibitem[Williams et al.(1991)]{1991ApJ...376..721W} Williams, R.~E., Hamuy, M., Phillips, M.~M., et al.\ 1991, ApJ, 376, 721 


\end{thebibliography}
\end{document}